# Invariant quadratic operators associated with Linear Canonical Transformations and their eigenstates

**Ravo Tokiniaina Ranaivoson[1], Raoelina Andriambololona[2], Hanitriarivo Rakotoson[3] Manjakamanana Rivo Herivola Ravelonjato[4]**

*tokhiniaina@gmail.com[1], raoelina.andriambololona@gmail.com[2], jacquelineraoelina@hotmail.com[2],raoelinasp@yahoo.fr[2], infotsara@gmail.com[3] manjakamanana@yahoo.fr[4]*

*Information Technology and Theoretical Physics Department[1,2,3,4]*
*Institut National des Sciences et Techniques Nucléaires (INSTN- Madagascar)*
BP 3907 Antananarivo101, Madagascar, *instn@moov.mg*

TWAS *Madagascar Chapter, Malagasy Academy[1,2,3]*
BP 4279 Antananarivo 101, Madagascar

*Faculty of Sciences – University of Antananarivo[3]*

***Abstract:*** The main purpose of this work is to identify invariant quadratic operators associated with Linear Canonical Transformations (LCTs) which could play important roles in physics. LCTs are considered in many fields. In quantum theory, they can be identified with linear transformations which keep invariant the Canonical Commutation Relations (CCRs). In this work, LCTs corresponding to a general pseudo-Euclidian space are considered and related to a phase space representation of quantum theory. Explicit calculations are firstly performed for the monodimensional case to identify the corresponding LCT-invariant quadratic operators then multidimensional generalizations of the obtained results are deduced. The eigenstates of these operators are also identified. A first kind of LCT-invariant operator is a second order polynomial of the coordinates and momenta operators and is a generalization of reduced momentum dispersion operator. The coefficients of this polynomial depend on the mean values and the statistical variances-covariances of the coordinates and momenta operators themselves. It is shown that these statistical variances-covariances can be related with thermodynamic variables. Two other LCT-invariant quadratic operators, which can be considered as the number operators of some quasipartciles, are also identified: the first one is a number operator of bosonic type quasiparticles and the second one corresponds to fermionic type. This fermionic LCT-invariant quadratic operator is directly related to a spin representation of LCTs. It is shown explicitly, in the case of a pentadimensional theory, that the eigenstates of this operator can be considered as basic quantum states of elementary fermions. A classification of the fundamental fermions, compatible with the Standard model of particle physics, is established from a classification of these states.

## 1-Introduction

Linear Canonical Transformations (LCTs) can be considered as generalization of some useful integral transformations like Fourier and Fractional Fourier Transforms. They are studied and used in many areas [1-11]. In quantum theory, they can be identified as the linear transformations which keep invariant the Canonical Commutation Relations (CCRs) defining the coordinates and momenta operators [9-11]. In the references [10-11], it was shown that a covariance principle related with LCTs may play an important role in physics and in the establishment of a unified theory of fundamental interactions. In the present work, our main objective is to identify some quadratic operators that are invariant under the action of LCTs and to highlight their potential importance in physics.

Some operators related to LCTs and their representations were already considered by various authors [5-6]. However, the operators that we consider through this work are new ones or can be considered as generalization of operators introduced in our previous works [9-11].

In the next section (section 2) we consider a study on a relation that exists between the simple theory of Harmonic oscillator and monodimensional LCTs. The goal of this section is to highlight some interesting basic facts and results that will be extended and generalized in the identification of the invariant quadratic operators associated to general LCTs and a phase space representation of quantum theory. These extensions and generalizations are performed in the sections 3 and 5. The statistical variance-covariances associated to the coordinate and momenta operators have an important place in the formalism that is established. In the section 4, it is shown that these statistical parameters can be practically related to thermodynamic variables. The main results obtained through this work are listed in the section 6 and conclusions are given in the section 7. The notation system which is considered is based on the reference [12]. Boldfaced letter like $\boldsymbol{p}$ are mainly used for quantum operator.

## 2- Quantum Linear Harmonic Oscillator and particular LCTs

### 2.1 Hamiltonian as invariant quadratic operator associated to particular LCTs

An example of a well-known quadratic operator that can be considered as invariant under the action of some particular monodimensional LCTs, in the framework of non-relativistic quantum theory, is the Hamiltonian operator $\boldsymbol{H}$ of a monodimensional harmonic oscillator. For a quantum harmonic oscillator with a mass $m$, angular frequency $\omega$, momentum operator $\boldsymbol{p}$ and coordinate operator $\boldsymbol{x}$ , the expression of this hamiltonian operator is

$$\boldsymbol{H} = \frac{(\boldsymbol{p})^2}{2m} + \frac{1}{2}m\omega^2(\boldsymbol{x})^2 \tag{1}$$

we may consider the set of linear transformations of the form

$$\begin{cases} \boldsymbol{p}' = \mathbb{a}\boldsymbol{p} + \mathbb{b}\boldsymbol{x} \\ \boldsymbol{x}' = \mathbb{c}\boldsymbol{p} + \mathbb{d}\boldsymbol{x} \end{cases} \quad \text{with} \begin{pmatrix} \mathbb{a} & \mathbb{c} \\ \mathbb{b} & \mathbb{d} \end{pmatrix} = \begin{pmatrix} cos\theta & -\dfrac{sin\theta}{m\omega} \\ m\omega sin\theta & cos\theta \end{pmatrix} \tag{2}$$

The relations (2) correspond to a LCT i.e. linear transformation which leaves invariant the CCR. We have the relations ($\hbar$ is the reduced Planck constant)

$$\mathbb{a}\mathbb{d} - \mathbb{b}\mathbb{c} = 1 \Leftrightarrow [\boldsymbol{p}', \boldsymbol{x}'] = [\boldsymbol{p}, \boldsymbol{x}] = -i\hbar \tag{3}$$

the Hamiltonian operator $\boldsymbol{H}$ in (1) is invariant under the action of the particular LCTs in (2)

$$\boldsymbol{H}' = \frac{(\boldsymbol{p}')^2}{2m} + \frac{1}{2}m\omega^2(\boldsymbol{x}')^2 = \frac{(\boldsymbol{p})^2}{2m} + \frac{1}{2}m\omega^2(\boldsymbol{x})^2 = \boldsymbol{H} \tag{4}$$

It will be noticed through the next sections that the expression of some general invariant quadratic operators associated to LCTs has a similarity with this Hamiltonian operator. Because of this similarity, the formalism that is considered in the identification of the eigenstates of theses operators shares analogy to the well-known formalism associated with the theory of harmonic oscillator. These eigenstates present some similarities with what are called coherent states, generalized coherent states and squeezed states in the literature [13-16]. It follows that the formalism considered in the present work can also be considered as an extension and generalization of the theory of quantum harmonic oscillator with an establishment of a link between it and a general theory of linear canonical transformations, phase space representation of quantum theory, thermodynamic and particle physics.

### 2.2 Statistical parameters, dispersion operators and ladder operators

It is well known that the eigenvalues equations of the Hamiltonian $\boldsymbol{H}$ in (1) is of the form

$$\boldsymbol{H}|n\rangle = (n + \frac{1}{2})\hbar\omega|n\rangle \tag{5}$$

with $n$ a non-negative integer. The wavefunctions corresponding to a state $|n\rangle$ in coordinate representations is of the form

$$\langle x|n\rangle = \frac{1}{\sqrt{2^n n!}}(\frac{m\omega}{\pi\hbar})^{1/4} H_n(\sqrt{\frac{m\omega}{\hbar}}x) e^{-\frac{m\omega}{2\hbar}x^2} \tag{6}$$

It can be easily deduced that the statistical mean values $\langle p\rangle$ and $\langle x\rangle$ of the momentum and coordinate operators associated to any eigenstate $|n\rangle$ of the Hamiltonian and the statistical variance $\mathcal{B}$ and $\mathcal{A}$ of these operators associated to the ground state $|0\rangle$ are respectively

$$\begin{cases} \langle p\rangle = \langle n|\boldsymbol{p}|n\rangle = 0 \quad \langle x\rangle = \langle n|\boldsymbol{x}|n\rangle = 0 \\ \mathcal{B} = \langle 0|(\boldsymbol{p} - \langle p\rangle)^2|0\rangle = \frac{m\hbar\omega}{2} = \frac{\hbar^2}{4\mathcal{A}} \\ \mathcal{A} = \langle 0|(\boldsymbol{x} - \langle x\rangle)^2|0\rangle = \frac{\hbar}{2m\omega} = \frac{\hbar^2}{4\mathcal{B}} \end{cases} \tag{7}$$

Using these statistical parameters, the expression of the Hamiltonian (4) can be written in the form

$$\boldsymbol{H} = \frac{(\boldsymbol{p})^2}{2m} + \frac{1}{2}m\omega^2(\boldsymbol{x})^2 = \frac{1}{2m}[(\boldsymbol{p} - \langle p\rangle)^2 + 4\mathcal{B}(\boldsymbol{x} - \langle x\rangle)^2] = \frac{\beth^{+}}{m} \tag{8}$$

in which $\beth^{+}$ is the momentum dispersion operator considered in the reference [9]. Here we have $\langle p\rangle = 0$ and $\langle x\rangle = 0$ according to (7) but they are introduced explicitly in the relation (8) to highlight the relation between the hamiltonian operator and the momentum dispersion operator that will be generalized in the next section.

Using again the relations in (7), the wavefunctions in the relation (6) can also be written in the form

$$\langle x|n,\mathcal{B}\rangle = \left(\frac{2\mathcal{B}}{\pi\hbar^2}\right)^{1/4}\frac{H_n\left(\frac{\sqrt{2\mathcal{B}}}{\hbar}x\right)}{\sqrt{2^n n!}}e^{-\frac{\mathcal{B}}{\hbar^2}x^2} = \left(\frac{1}{2\pi\mathcal{A}}\right)^{1/4}\frac{H_n\left(\frac{1}{\sqrt{2\mathcal{A}}}x\right)}{\sqrt{2^n n!}}e^{-\frac{1}{4\mathcal{A}}x^2} \qquad (9)$$

and the eigenvalue equation (5) of the Hamiltonian operator itself can be written in the form

$$\boldsymbol{H}|n,\mathcal{B}\rangle = \frac{ב^+}{m}|n,\mathcal{B}\rangle = (2n+1)\frac{\mathcal{B}}{m}|n,\mathcal{B}\rangle \qquad (10)$$

The change of notation from $|n\rangle$ to $|n,\mathcal{B}\rangle$ is introduced to highlight the fact that in reality there is also a dependance on $\mathcal{B}$.

Dirac in his theory of harmonic oscillator has introduced the ladder operators (annihilation and creation operators). Using the statistical parameters considered previously, the expression of these ladder operators (denoted $\boldsymbol{\mathcal{z}}$ and $\boldsymbol{\mathcal{z}}^\dagger$) can be written in the forms

$$\begin{cases} \boldsymbol{\mathcal{z}} = \sqrt{\frac{m\omega}{2\hbar}}\left(\boldsymbol{x} + \frac{i}{m\omega}\boldsymbol{p}\right) = \frac{i}{2\sqrt{\mathcal{B}}}\left(\boldsymbol{p} - \frac{2i}{\hbar}\mathcal{B}\boldsymbol{x}\right) = \frac{i\boldsymbol{z}}{2\sqrt{\mathcal{B}}} \\ \boldsymbol{\mathcal{z}}^\dagger = \sqrt{\frac{m\omega}{2\hbar}}\left(\boldsymbol{x} - \frac{i}{m\omega}\boldsymbol{p}\right) = \frac{-i}{2\sqrt{\mathcal{B}}}\left(\boldsymbol{p} + \frac{2i}{\hbar}\mathcal{B}\boldsymbol{x}\right) = \frac{-i\boldsymbol{z}^\dagger}{2\sqrt{\mathcal{B}}} \end{cases} \quad [\boldsymbol{\mathcal{z}},\boldsymbol{\mathcal{z}}^\dagger] = 1 \qquad (11)$$

The operators $\boldsymbol{z} = \boldsymbol{p} + \frac{2i}{\hbar}\mathcal{B}\boldsymbol{x}$ is introduced in the relation (11) because the generalization of it plays an important role in the multidimensional generalization and in the construction of the phase space representation that are considered in the next section. This operator is a particular monodimensional case of the operator $\boldsymbol{z}_\mu = \boldsymbol{p}_\mu + \frac{2i}{\hbar}\mathcal{B}_{\mu\nu}\boldsymbol{x}^\nu$introduced in the reference [10].
In term of the ladder operators $\boldsymbol{\mathcal{z}}$ and $\boldsymbol{\mathcal{z}}^\dagger$, we have for $\boldsymbol{H}$

$$\boldsymbol{H} = \frac{ב^+}{m} = \frac{1}{4m}(\boldsymbol{z}^\dagger\boldsymbol{z} + \boldsymbol{z}\boldsymbol{z}^\dagger) = \frac{\mathcal{B}}{m}(\boldsymbol{\mathcal{z}}^\dagger\boldsymbol{\mathcal{z}} + \boldsymbol{\mathcal{z}}\boldsymbol{\mathcal{z}}^\dagger) = \frac{\mathcal{B}}{m}(2\boldsymbol{\mathcal{z}}^\dagger\boldsymbol{\mathcal{z}} + 1) \qquad (12)$$

The eigenvalue equation (9) can be itself deduced from the relation (12) and the commutation relation in (11). We have also the well-known relations

$$\begin{cases} \boldsymbol{\mathcal{z}}|n,\mathcal{B}\rangle = \sqrt{n}|n-1,\mathcal{B}\rangle \\ \boldsymbol{\mathcal{z}}^\dagger|n,\mathcal{B}\rangle = \sqrt{n+1}|n+1,\mathcal{B}\rangle \end{cases} \qquad (13)$$

which justify the name ladder operators for $\boldsymbol{\mathcal{z}}$ and $\boldsymbol{\mathcal{z}}^\dagger$. Under the action of the LCT (2), the law of transformation of the ladder operators $\boldsymbol{\mathcal{z}}$ and $\boldsymbol{\mathcal{z}}^\dagger$are unitary transformations

$$\begin{cases} \boldsymbol{\mathcal{z}}' = e^{i\theta}\boldsymbol{\mathcal{z}} \\ \boldsymbol{\mathcal{z}}'^\dagger = e^{-i\theta}\boldsymbol{\mathcal{z}}^\dagger \end{cases} \qquad (14)$$

The unitary transformation (14) is also known as a Bogolioubov transformation . The LCT-invariance of the Hamiltonian can also be seen through the relation (12) and (14).

### 2.3 Harmonic oscillator with nonzero values of $\langle p\rangle$ and $\langle x\rangle$

Instead of an oscillator with the mean values $\langle p\rangle = 0$ and $\langle x\rangle = 0$ considered previously, we may consider an oscillator that may have non-null momentum and coordinate mean values: $\langle p\rangle \neq 0$ and $\langle x\rangle \neq 0$. Instead of the relation (8), we should have for the basic wavefunctions of this kind of harmonic oscillator, in coordinate representation,

$$\langle x|n,\langle p\rangle,\langle x\rangle,\mathcal{B}\rangle = (\frac{2\mathcal{B}}{\pi\hbar^2})^{1/4}\frac{H_n(\frac{\sqrt{2\mathcal{B}}}{\hbar}(x-\langle x\rangle))}{\sqrt{2^n n!}}e^{-\frac{\mathcal{B}}{\hbar^2}(x-\langle x\rangle)^2+\frac{i}{\hbar}\langle p\rangle x} \tag{15}$$

The operator which admit the states $|n,\langle p\rangle,\langle x\rangle,\mathcal{B}\rangle$corresponding to these wavefunctions as eigenstates is the momentum dispersion operator

$$\beth^+ = \frac{1}{2}[(\boldsymbol{p}-\langle p\rangle)^2 + 4\mathcal{B}(\boldsymbol{x}-\langle x\rangle)^2] \tag{16}$$

with the eigenvalue equation

$$\beth^+|n,\langle p\rangle,\langle x\rangle,\mathcal{B}\rangle = (2n+1)\mathcal{B}|n,\langle p\rangle,\langle x\rangle,\mathcal{B}\rangle \tag{17}$$

The expression of the new hamiltonian $\boldsymbol{H}$ which generalizes (8) for $\langle p\rangle \neq 0$ and $\langle x\rangle \neq 0$ should be

$$\boldsymbol{H} = \frac{\langle p\rangle^2}{2m} + \frac{\beth^+}{m} = \frac{\langle p\rangle^2}{2m} + \frac{1}{2m}[(\boldsymbol{p}-\langle p\rangle)^2 + 4\mathcal{B}(\boldsymbol{x}-\langle x\rangle)^2] \tag{18}$$

The term $\frac{\langle p\rangle^2}{2m}$ is the mean values of the kinetic energy and the term $\frac{\beth^+}{m}$ describes the part of energy corresponding to the quantum fluctuation-oscillation around the means values $\langle p\rangle$ and $\langle x\rangle$. When $\langle p\rangle = 0$ and $\langle x\rangle = 0$, the relation (18) is reduced as expected to the relation (8). The eigenvalue equation of the hamiltonian $\boldsymbol{H}$ itself can be deduced easily from (19)

$$\boldsymbol{H}|n,\langle p\rangle,\langle x\rangle,\mathcal{B}\rangle = [\frac{\langle p\rangle^2}{2m} + (2n+1)\frac{\mathcal{B}}{m}]|n,\langle p\rangle,\langle x\rangle,\mathcal{B}\rangle \tag{19}$$

Using the relation (15), it can be checked that a state $|0,\langle p\rangle,\langle x\rangle,\mathcal{B}\rangle$ $(n=0)$ is an eigenstate of the $\boldsymbol{z} = \boldsymbol{p} + \frac{2i}{\hbar}\mathcal{B}\boldsymbol{x}$ with the eigenvalue $\langle z\rangle = \langle n,\langle p\rangle,\langle x\rangle,\mathcal{B}|\boldsymbol{z}|n,\langle p\rangle,\langle x\rangle,\mathcal{B}\rangle = \langle p\rangle + \frac{2i}{\hbar}\mathcal{B}\langle x\rangle$

$$\boldsymbol{z}|0,\langle p\rangle,\langle x\rangle,\mathcal{B}\rangle = [\langle p\rangle + \frac{2i}{\hbar}\mathcal{B}\langle x\rangle]|0,\langle p\rangle,\langle x\rangle,\mathcal{B}\rangle = \langle z\rangle|0,\langle p\rangle,\langle x\rangle,\mathcal{B}\rangle \tag{20}$$

It follows that we may use the simpler notations $|\langle z\rangle\rangle = |0,\langle p\rangle,\langle x\rangle,\mathcal{B}\rangle$ and $|n,\langle p\rangle,\langle x\rangle,\mathcal{B}\rangle = |n,\langle z\rangle\rangle$ (as in the reference [10]).

As a generalization of the ladder operators $\boldsymbol{ƶ}$ and $\boldsymbol{ƶ}^\dagger$ in the relations (11) and (13), we have

$$\begin{cases} \boldsymbol{ƶ} = \dfrac{i(\boldsymbol{z}-\langle z\rangle)}{2\sqrt{\mathcal{B}}} = \dfrac{i}{2\sqrt{\mathcal{B}}}[(\boldsymbol{p}-\langle p\rangle) - \dfrac{2i}{\hbar}\mathcal{B}(\boldsymbol{x}-\langle x\rangle)] \\ \boldsymbol{ƶ}^\dagger = \dfrac{i(\boldsymbol{z}-\langle z\rangle)}{2\sqrt{\mathcal{B}}} = \dfrac{i}{2\sqrt{\mathcal{B}}}[(\boldsymbol{p}-\langle p\rangle) - \dfrac{2i}{\hbar}\mathcal{B}(\boldsymbol{x}-\langle x\rangle)] \end{cases} \quad [\boldsymbol{ƶ},\boldsymbol{ƶ}^\dagger] = 1 \tag{21}$$

$$\begin{cases} \boldsymbol{z}|n,\langle z\rangle\rangle = \sqrt{n}|n-1,\langle z\rangle\rangle \\ \boldsymbol{z}^{\dagger}|n,\langle z\rangle\rangle = \sqrt{n+1}|n+1,\langle z\rangle\rangle \end{cases} \tag{22}$$

and we have also as generalization of (12)

$$\boldsymbol{H} = \frac{\langle p\rangle^2}{2m} + \frac{\beth^+}{m} = \frac{\langle p\rangle^2}{2m} + \frac{1}{m}(\boldsymbol{z}^{\dagger}\boldsymbol{z} + \boldsymbol{z}\boldsymbol{z}^{\dagger}) = \frac{\langle p\rangle^2}{2m} + \frac{\mathcal{B}}{m}(2\boldsymbol{z}^{\dagger}\boldsymbol{z} + 1) \tag{23}$$

It is straightforward to remark that the relations (21), (22) and (23) correspond to the eigenvalues equation (19).

## 3- Phase space representation and LCTs: non-relativistic monodimensional case

### 3.1 Analogy between a non-relativistic quantum particle and a harmonic oscillator

In the framework of classical mechanics, the simplest phase space is the phase space of a Newtonian material point performing a monodimesional motion. This motion, is classically, completely determined by the instantaneous values of its momentum $p$ and coordinate $x$. The corresponding phase space is the set $\{(p,x)\}$ of all possible values of $(p,x)$ and the classical state is then completely defined by a point (i.e. an element) of this classical phase space.

In the framework of quantum mechanics, the uncertainty principle tells that it is impossible to have simultaneously an exact values of $p$ and $x$. It follows that the state of a particle cannot be represented with a couple $(p,x)$ and then the concept of phase space as defined from classical mechanics lead to an ambiguity.

In this section, our goal is to describe a formalism that can lead to an acceptable definition of a quantum phase space. It is achieved through the introduction of a phase space representation of quantum states and operators using some results based on the theory of linear harmonic oscillator and its relation with LCTs as described through the previous section.

In the framework of quantum mechanics, the particle is described with a state $|\psi\rangle$ to which correspond a coordinate wavefunction $\psi(x)$ and a momentum wavefunction $\tilde{\psi}(p)$. The wavefunctions $\psi(x)$ and $\tilde{\psi}(p)$ are linked by a Fourier transform. Let us denote respectively $\langle p\rangle, \langle x\rangle, \mathcal{B}$ and $\mathcal{A}$ the mean values and statistical variance of the momentum operator $\boldsymbol{p}$ and coordinate operator $\boldsymbol{x}$ of the particle.

$$\begin{cases} \langle\psi|\boldsymbol{p}|\psi\rangle = \langle p\rangle \quad \langle\psi|(\boldsymbol{p}-\langle p\rangle)^2|\psi\rangle = \mathcal{B} \\ \langle\psi|\boldsymbol{x}|\psi\rangle = \langle x\rangle \quad \langle\psi|(\boldsymbol{x}-\langle x\rangle)^2|\psi\rangle = \mathcal{A} \end{cases} \tag{24}$$

we have the uncertainty relation

$$\sqrt{\mathcal{A}}\sqrt{\mathcal{B}} \geq \frac{\hbar}{2} \tag{25}$$

In the ordinary formulation of non-relativistic quantum mechanics, the Hamiltonian $\boldsymbol{H}$ of an ideal free particle (considered as a particle with $\mathcal{B} = 0$ and then $\mathcal{A} \to +\infty$) of mass $m$ is

$$\boldsymbol{H} = \frac{(\boldsymbol{p})^2}{2m} \tag{26}$$

$\boldsymbol{H}$ and $\boldsymbol{p}$ has the same eigenstates $|p\rangle$ which correspond, in coordinate representation, to wavefunctions that are plane waves.

$$\langle x|p\rangle = \frac{1}{\sqrt{2\pi\hbar}} e^{-\frac{i}{\hbar}px} \tag{27}$$

The eigenvalues equation are

$$\begin{cases} \boldsymbol{p}|p\rangle = p|p\rangle \\ \boldsymbol{H}|p\rangle = \dfrac{(p)^2}{2m}|p\rangle \end{cases} \tag{28}$$

But to be more realistic and to have a practical compatibility with the uncertainty relation (25) which take into account "quantum fluctuation", we should consider that we have $\mathcal{B} > 0$ which correspond, in coordinates representations, to wave packets. In that case, we should find a general expression of the Hamiltonian that leads to the particular case described by the relations (26), (27) and (28) in the ideal limit $\mathcal{B} \to 0$. We may suppose that the energy of a quantum particle of mass $m$ with a momentum and coordinate mean values $\langle p\rangle$ and $\langle x\rangle$ is composed of two parts: a mean kinetic energy and an energy which corresponds to quantum fluctuation of momentum. In other words, we may postulate that the hamiltonian of a real quantum particle is exactly similar to the hamiltonian (23) of an harmonic oscillator with nonzero momentum and coordinate mean values

$$\boldsymbol{H} = \frac{\langle p\rangle^2}{2m} + \frac{\beth^{+}}{m} = \frac{\langle p\rangle^2}{2m} + \frac{1}{m}(\boldsymbol{z}^{\dagger}\boldsymbol{z} + \boldsymbol{z}\boldsymbol{z}^{\dagger}) = \frac{\langle p\rangle^2}{2m} + \frac{\mathcal{B}}{m}(2\boldsymbol{z}^{\dagger}\boldsymbol{z} + 1) \tag{29}$$

$\frac{\langle p\rangle^2}{2m}$ is the mean kinetic energy and $\frac{\beth^{+}}{m}$ describes the part of the energy corresponding to quantum fluctuation. In the limit $\mathcal{B} \to 0$(quantum fluctuation of momentum are neglected), one has $\langle p\rangle \longrightarrow p$ and one obtain the case of an ideal free particle.

Following the relation (29), we have an exact analogy between the motion of a quantum particle and an harmonic oscillator with nonzero momentum and coordinate mean values described previously. We have, in particular, for the eigenvalue equation of the hamiltonian

$$\boldsymbol{H}|n,\langle z\rangle\rangle = (\frac{\langle p\rangle^2}{2m} + \frac{\beth^{+}}{m})|n,\langle z\rangle\rangle = [\frac{\langle p\rangle^2}{2m} + (2n+1)\frac{\mathcal{B}}{m}]|n,\langle z\rangle\rangle \tag{30}$$

The set $\{|n,\langle z\rangle\rangle\}$ of the eigenstates of the Hamiltonian is an orthonormal basis of the states space of the particle. For any state $|\psi\rangle$ of the particle, we have the decomposition

$$|\psi\rangle = \sum_n |n,\langle z\rangle\rangle\langle n,\langle z\rangle|\psi\rangle = \sum_n \Psi^n(\langle p\rangle,\langle x\rangle,\mathcal{B})|n,\langle z\rangle\rangle \tag{31}$$

with the wavefunction $\Psi^n(\langle p\rangle,\langle x\rangle,\mathcal{B}) = \langle n,\langle z\rangle|\psi\rangle$. $|\Psi^n(\langle p\rangle,\langle x\rangle,\mathcal{B})|^2 = |\langle n,\langle z\rangle|\psi\rangle|^2$ is the probability to find the particle in the state $|n,\langle z\rangle\rangle$ when the state is $|\psi\rangle$.

### 3.2 Phase space representation and quantum phase space

The eigenstates of the operator $\boldsymbol{z}$ corresponding to the quantum particle are the states $|\langle z\rangle\rangle = |0,\langle p\rangle,\langle x\rangle,\mathcal{B}\rangle$ with an eigenvalue equation exactly similar to (22)

$$\boldsymbol{z}|\langle z\rangle\rangle = \langle z\rangle|\langle z\rangle\rangle = (\langle p\rangle + \frac{2i}{\hbar}\mathcal{B}\langle x\rangle)|\langle z\rangle\rangle \tag{32}$$

A state $|\langle z\rangle\rangle$ can be explicitly and completely characterized by the corresponding wavefunction $\langle x|\langle z\rangle\rangle$ in coordinate representation. It can be deduced as a particular case of the relation (15) for $n = 0$

$$\langle x|\langle z\rangle\rangle = \langle x|0,\langle p\rangle,\langle x\rangle,\mathcal{B}\rangle = (\frac{2\mathcal{B}}{\pi\hbar^2})^{1/4} e^{-\frac{\mathcal{B}}{\hbar^2}(x-\langle x\rangle)^2+\frac{i}{\hbar}\langle p\rangle x} = (\frac{1}{2\pi\mathcal{A}})^{1/4} e^{-\frac{1}{4\mathcal{A}}(x-\langle x\rangle)^2+\frac{i}{\hbar}\langle p\rangle x} \tag{33}$$

with $\mathcal{A} = \frac{(\hbar)^2}{4\mathcal{B}}$ the coordinate statistical variance when the particle is in a state $|\langle z\rangle\rangle$.The set $\{|\langle z\rangle\rangle\}$ of the eigenstates of $\boldsymbol{z}$ is not an orthonormal basis ($\boldsymbol{z}$ is not hermitian). Explicitly, it can be checked we have the scalar product

$$\langle\langle z\rangle|\langle z\rangle'\rangle = \int \langle\langle z\rangle|x\rangle\langle x|\langle z\rangle'\rangle\, dx = e^{-\frac{(\langle p\rangle-\langle p'\rangle)^2}{8\mathcal{B}} - \frac{(\langle x\rangle-\langle x'\rangle)^2}{8\mathcal{A}} - i\frac{(\langle p\rangle-\langle p'\rangle)(\langle x\rangle+\langle x'\rangle)}{2\hbar}} \tag{34}$$

However, we have the dyadic decomposition ($h = 2\pi\hbar$ is the Planck constant)

$$\iint |\langle z\rangle\rangle\langle\langle z\rangle| \frac{d\langle p\rangle d\langle x\rangle}{h} = \boldsymbol{I} \tag{35}$$

in which $\boldsymbol{I}$ is the identity operator. It follows from the relation (35) that for any possible quantum state $|\psi\rangle$ of the particle and for any operator $\boldsymbol{A}$, we have the decompositions

$$|\psi\rangle = \boldsymbol{I}|\psi\rangle = \iint |\langle z\rangle\rangle\langle\langle z\rangle|\psi\rangle \frac{d\langle p\rangle d\langle x\rangle}{h} = \iint \Psi^0(\langle p\rangle,\langle x\rangle,\mathcal{B})|\langle z\rangle\rangle \frac{d\langle p\rangle d\langle x\rangle}{h} \tag{36}$$

$$\boldsymbol{A} = \boldsymbol{IAI} = \iint \iint |\langle z\rangle\rangle\langle\langle z\rangle|\boldsymbol{A}|\langle z\rangle'\rangle \langle\langle z\rangle'| \frac{d\langle p\rangle d\langle x\rangle}{h} \frac{d\langle p\rangle' d\langle x\rangle'}{h} \tag{37}$$

The relations (36) and (37) can be considered as defining "phase spaces representations" of a state $|\psi\rangle$ and of a quantum operator $\boldsymbol{A}$ : they define a "phase space representation of quantum mechanics". In this point of view, a "quantum phase space" is, according to relations (35), (36) and (37), to be identified with the set $\{\langle p\rangle,\langle x\rangle\}$ for a given value of $\mathcal{B}$.

We have for the wavefunction $\Psi^0(\langle p\rangle,\langle x\rangle,\mathcal{B}) = \langle\langle z\rangle|\psi\rangle$ in (38) the normalization relation

$$\iint |\langle\langle z\rangle|\psi\rangle|^2 \frac{d\langle p\rangle d\langle x\rangle}{h} = 1 \Leftrightarrow \iint |\langle\langle z\rangle|\psi\rangle|^2\, d\langle p\rangle d\langle x\rangle = h \tag{38}$$

It can be remarked that the presence of the Planck constant $h = 2\pi\hbar$ in the relations (35), (36), (37) and (38) may be related to the important fact that in statistical mechanics the surface of a "phase space elementary cell" should be taken to be equal to $h$ . The normalization relation (38) can be considered as saying that $\Psi^0(\langle p\rangle,\langle x\rangle,\mathcal{B}) = |\langle\langle z\rangle|\psi\rangle|^2$corresponds to a description of the "spreadness" of the particle microstate $|\psi\rangle$ on the quantum phase space.

**Remark:** we may consider the following operators denoted $\boldsymbol{\mathscr{p}}, \boldsymbol{\mathscr{x}}$ and $\beth^+$ that was already introduced in the references [9-10]. They are called respectively reduced momentum operator, reduced coordinate operator and reduced momentum dispersion operator

$$\begin{cases} \boldsymbol{\mathscr{p}} = \frac{1}{\sqrt{2}}(\boldsymbol{\mathscr{z}}^\dagger + \boldsymbol{\mathscr{z}}) = \frac{i}{\sqrt{2\mathcal{B}}}(\boldsymbol{p} - \langle p \rangle) = \frac{i}{\hbar}\sqrt{2\mathcal{A}}(\boldsymbol{p} - \langle p \rangle) \\ \boldsymbol{\mathscr{x}} = \frac{i}{\sqrt{2}}(\boldsymbol{\mathscr{z}}^\dagger - \boldsymbol{\mathscr{z}}) = \frac{1}{i\sqrt{2\mathcal{A}}}(\boldsymbol{x} - \langle x \rangle) = \frac{1}{i\hbar}\sqrt{2\mathcal{B}}(\boldsymbol{x} - \langle x \rangle) \\ \beth^+ = \frac{\beth^+}{4\mathcal{B}} = \frac{1}{4}(\boldsymbol{\mathscr{z}}^\dagger\boldsymbol{\mathscr{z}} + \boldsymbol{\mathscr{z}}\boldsymbol{\mathscr{z}}^\dagger) = -\frac{1}{4}[(\boldsymbol{\mathscr{p}})^2 + (\boldsymbol{\mathscr{x}})^2] \end{cases} \quad (39)$$

As shown in the reference [10], the generalization of these operators plays an important role in the study of general multidimensional LCTs. The generalization of the operator $\beth^+$, in particular, is an invariant quadratic operator associated to LCTs.

### 3.3 Invariant quadratic operator associated to general monodimensional LCTs

The LCTs considered in the relation (2) is a particular case of monodimensional LCTs. general ones satisfy the relation (algebraic covariant and contravariant index are introduced)

$$\begin{cases} \boldsymbol{p}'_1 = \mathbb{a}_1^1\boldsymbol{p_1} + \mathbb{b}_1^1\boldsymbol{x_1} \\ \boldsymbol{x}'_1 = \mathbb{c}_1^1\boldsymbol{p_1} + \mathbb{d}_1^1\boldsymbol{x_1} \\ [\boldsymbol{p}'_1, \boldsymbol{x}'_1] = [\boldsymbol{p_1}, \boldsymbol{x_1}] = -i\hbar \end{cases} \Leftrightarrow \mathbb{a}_1^1\mathbb{d}_1^1 - \mathbb{b}_1^1\mathbb{c}_1^1 = 1 \Leftrightarrow \begin{pmatrix} \mathbb{a}_1^1 & \mathbb{c}_1^1 \\ \mathbb{b}_1^1 & \mathbb{d}_1^1 \end{pmatrix} \in SL(2) \quad (40)$$

in which $SL(2)$is the Special Linear group.

The identification of the commutation relation $[\boldsymbol{p}'_1, \boldsymbol{x}'_1] = [\boldsymbol{p_1}, \boldsymbol{x_1}] = -i\hbar$ as a monodimensional particular case of the general multidimensional form $[\boldsymbol{p}_\mu, \boldsymbol{x}_\nu] = i\hbar\eta_{\mu\nu}$ considered in the reference [10] permits to identify "the metric" $\eta = \eta_{11} = -1$.

As shown in [10], the general relation for a $2D \times 2D$ matrix $\begin{pmatrix} \mathbb{a} & \mathbb{c} \\ \mathbb{b} & \mathbb{d} \end{pmatrix}$ corresponding to an LCT in a $D$-dimensional pseudo-Euclidian space with a metric $\eta$ with signature $(D_+, D_-)$is

$$\begin{pmatrix} \mathbb{a} & \mathbb{c} \\ \mathbb{b} & \mathbb{d} \end{pmatrix}^T \begin{pmatrix} 0 & \eta \\ -\eta & 0 \end{pmatrix} \begin{pmatrix} \mathbb{a} & \mathbb{c} \\ \mathbb{b} & \mathbb{d} \end{pmatrix} = \begin{pmatrix} 0 & \eta \\ -\eta & 0 \end{pmatrix} \quad (41)$$

and $\begin{pmatrix} \mathbb{a} & \mathbb{c} \\ \mathbb{b} & \mathbb{d} \end{pmatrix}$ belongs then to a symplectic group $Sp(2D_+, 2D_-)$. For the space associated to the monodimensional LCT in (40), the signature is (0,1) i.e $\eta = \eta_{11} = -1$ and we have the explicit equivalence

$$\begin{pmatrix} \mathbb{a}_1^1 & \mathbb{c}_1^1 \\ \mathbb{b}_1^1 & \mathbb{d}_1^1 \end{pmatrix}^T \begin{pmatrix} 0 & -1 \\ 1 & 0 \end{pmatrix} \begin{pmatrix} \mathbb{a}_1^1 & \mathbb{c}_1^1 \\ \mathbb{b}_1^1 & \mathbb{d}_1^1 \end{pmatrix} = \begin{pmatrix} 0 & -1 \\ 1 & 0 \end{pmatrix} \Leftrightarrow \mathbb{a}_1^1\mathbb{d}_1^1 - \mathbb{b}_1^1\mathbb{c}_1^1 = 1 \quad (42)$$

The relation (42) corresponds to the group isomorphism $Sp(0,2) \cong SL(2)$.

The LCT (42) can be considered as a transformation which affect the momentum and coordinate operators (and any operator depending on them) but doesn't affect the quantum state of a particle : this is the point of view considered in the references [10-11] in which a general multidimensional LCT is considered as describing a change of observational reference frame.
From now, for sake of simplicity, we will use the natural system unit in which one takes for the reduced Planck constant $\hbar = 1$. Let $|\psi\rangle$ be the state of a particle. We may introduce the following mean values and statistical variance-covariance

$$\begin{cases} \langle x_1\rangle = \langle\psi|\boldsymbol{x_1}|\psi\rangle = -\langle\psi|\boldsymbol{x^1}|\psi\rangle = -\langle x^1\rangle \\ \langle p_1\rangle = \langle\psi|\boldsymbol{p_1}|\psi\rangle = -\langle\psi|\boldsymbol{p^1}|\psi\rangle = -\langle p^1\rangle \\ \mathcal{X}_{11} = \langle\psi|(\boldsymbol{x_1} - \langle x_1\rangle)^2|\psi\rangle = -\mathcal{X}_1^1 = \mathcal{X}^{11} \\ \mathcal{P}_{11} = \langle\psi|(\boldsymbol{p_1} - \langle p_1\rangle)^2|\psi\rangle = -\mathcal{P}_1^1 = \mathcal{P}^{11} \\ \varrho_{11}^{\ltimes} = \langle\psi|(\boldsymbol{p_1} - \langle p_1\rangle)(\boldsymbol{x_1} - \langle x_1\rangle)|\psi\rangle \\ \varrho_{11}^{\rtimes} = \langle\psi|(\boldsymbol{x_1} - \langle x_1\rangle)(\boldsymbol{p_1} - \langle p_1\rangle)|\psi\rangle \\ \varrho_{11} = \frac{1}{2}(\varrho_{11}^{\ltimes} + \varrho_{11}^{\rtimes}) \end{cases} \quad (43)$$

$\varrho_{11}^{\ltimes}, \varrho_{11}^{\rtimes}$ and $\varrho_{11}$ can be both considered as some kind of momentum-coordinate statistical covariances (codispersions). But they are different because the operators $\boldsymbol{p}$ and $\boldsymbol{x}$ do not commute. However, it can be deduced from the commutation relation $[\boldsymbol{p_1}, \boldsymbol{x_1}] = -i$ that we have the relations

$$\varrho_{11}^{\ltimes} - \varrho_{11}^{\rtimes} = -i \Rightarrow \begin{cases} \varrho_{11}^{\ltimes} = \varrho_{11}^{\rtimes} - i = \varrho_{11} - \frac{i}{2} \\ \varrho_{11}^{\rtimes} = \varrho_{11}^{\ltimes} + i = \varrho_{11} + \frac{i}{2} \\ \varrho_{11} = \frac{1}{2}(\varrho_{11}^{\ltimes} + \varrho_{11}^{\rtimes}) = \varrho_{11}^{\ltimes} + \frac{i}{2} = \varrho_{11}^{\rtimes} - \frac{i}{2} \end{cases} \quad (44)$$

Under the action of the LCT (40), the mean values and statistical variance-covariance become

$$\begin{cases} \langle x_1'\rangle = \langle\psi|\boldsymbol{x_1'}|\psi\rangle = -\langle\psi|\boldsymbol{x^{1\prime}}|\psi\rangle = -\langle x'^1\rangle \\ \langle p_1'\rangle = \langle\psi|\boldsymbol{p_1'}|\psi\rangle = -\langle\psi|\boldsymbol{p^{1\prime}}|\psi\rangle = -\langle p'^1\rangle \\ \mathcal{X}_{11}' = \langle\psi|(\boldsymbol{x_1'} - \langle x_1'\rangle)^2|\psi\rangle = -\mathcal{X}_1'^1 = \mathcal{X}'^{11} \\ \mathcal{P}_{11}' = \langle\psi|(\boldsymbol{p_1'} - \langle p_1'\rangle)^2|\psi\rangle = -\mathcal{P}_1'^1 = \mathcal{P}'^{11} \\ \varrho_{11}^{\ltimes\prime} = \langle\psi|(\boldsymbol{p_1'} - \langle p_1'\rangle)(\boldsymbol{x_1'} - \langle x_1'\rangle)|\psi\rangle \\ \varrho_{11}^{\rtimes\prime} = \langle\psi|(\boldsymbol{x_1'} - \langle x_1'\rangle)(\boldsymbol{p_1'} - \langle p_1'\rangle)|\psi\rangle \\ \varrho_{11}' = \frac{1}{2}(\varrho_{11}^{\ltimes\prime} + \varrho_{11}^{\rtimes\prime}) \end{cases} \quad (45)$$

Using the relations (40), (43) and (45), we can deduce the law of transformations

$$(\langle p_1'\rangle \quad \langle x_1'\rangle) = (\langle p_1\rangle \quad \langle x_1\rangle)\begin{pmatrix} \mathbb{a}_1^1 & \mathbb{c}_1^1 \\ \mathbb{b}_1^1 & \mathbb{d}_1^1 \end{pmatrix} \quad (46)$$

$$\begin{pmatrix} \mathcal{P}'_{11} & \varrho'_{11} \\ \varrho'_{11} & \mathcal{X}'_{11} \end{pmatrix} = \begin{pmatrix} \mathbb{a}_1^1 & \mathbb{c}_1^1 \\ \mathbb{b}_1^1 & \mathbb{d}_1^1 \end{pmatrix}^T \begin{pmatrix} \mathcal{P}_{11} & \varrho_{11} \\ \varrho_{11} & \mathcal{X}_{11} \end{pmatrix} \begin{pmatrix} \mathbb{a}_1^1 & \mathbb{c}_1^1 \\ \mathbb{b}_1^1 & \mathbb{d}_1^1 \end{pmatrix} \tag{47}$$

Then, the following invariants can be deduced using the relation (42):

- An invariant scalar which is the determinant of the matrix $\begin{pmatrix} \mathcal{P}_{11} & \varrho_{11} \\ \varrho_{11} & \mathcal{X}_{11} \end{pmatrix}$ that we may call, following the reference [10], momentum –coordinate variance-covariance matrix

$$\begin{vmatrix} \mathcal{P}'_{11} & \varrho'_{11} \\ \varrho'_{11} & \mathcal{X}'_{11} \end{vmatrix} = \mathcal{P}'_{11}\mathcal{X}'_{11} - (\varrho'_{11})^2 = \mathcal{P}_{11}\mathcal{X}_{11} - (\varrho_{11})^2 = \begin{vmatrix} \mathcal{P}_{11} & \varrho_{11} \\ \varrho_{11} & \mathcal{X}_{11} \end{vmatrix} \tag{48}$$

- An invariant quadratic operator that we may denote $\beth^+$

$$\beth^+ = \frac{1}{2}\left((\boldsymbol{p_1} - \langle p_1\rangle) \quad (\boldsymbol{x_1} - \langle x_1\rangle)\right) \begin{pmatrix} \mathcal{X}_{11} & -\varrho_{11} \\ -\varrho_{11} & \mathcal{P}_{11} \end{pmatrix} \left((\boldsymbol{p_1} - \langle p_1\rangle) \quad (\boldsymbol{x_1} - \langle x_1\rangle)\right)^T \tag{49}$$

The notation $\beth^+$is chosen because this invariant quadratic operator in the relation (49) can be considered as a generalization of the reduced momentum dispersion operator in (39). In fact, in the limit $\varrho_{11} = 0$, it can be checked  that the two operators is exactly equal.

Using the coordinate representation, it can be shown that the eigenstate of the invariant quadratic operator $\beth^+$ in (49) corresponding to its lowest eigenvalue is a state, that we may denote $|\langle z_1\rangle\rangle$, corresponding to the coordinate wavefunction of the form

$$\langle x^1|\langle z_1\rangle\rangle = e^{iK}\left(\frac{1}{2\pi\mathcal{X}_{11}}\right)^{1/4} e^{-\mathcal{B}_{11}\left(x^1 - \langle x^1\rangle\right)^2 - i\langle p_1\rangle x^1} \tag{50}$$

in which $e^{iK}$ is an unitary complex number ($K$ is real number). $\langle p_1\rangle, \langle x^1\rangle$ and$\mathcal{X}_{11}$ are the mean values and statistical variances, as defined in (43), associated to the state$|\psi\rangle = |\langle z_1\rangle\rangle$ itself. $\mathcal{B}_{11}$ is a complex parameter given by the relation

$$\mathcal{B}_{11} = -i\frac{\varrho_{11}^{\ltimes}}{2\mathcal{X}_{11}} = \frac{1}{4\mathcal{X}_{11}} - i\frac{\varrho_{11}}{2\mathcal{X}_{11}} \tag{51}$$

With $\varrho_{11}^{\ltimes}$ and $\varrho_{11}$the momentum-coordinate covariances, as defined in (43), associated to  the state $|\psi\rangle = |\langle z_1\rangle\rangle$.If we denoted also $\mathcal{P}_{11}$ the momentum statistical variance for this state $|\psi\rangle = |\langle z_1\rangle\rangle$, it can be checked, using (50), that for this particular case the LCT-invariant scalar (48) is exactly given by  relation

$$\mathcal{P}'_{11}\mathcal{X}'_{11} - (\varrho'_{11})^2 = \mathcal{P}_{11}\mathcal{X}_{11} - (\varrho_{11})^2 = \frac{1}{4} \tag{52}$$

And this invariant scalar is exactly the lowest eigenvalue of the LCT-invariant quadratic operator $\beth^+$ in (49)  i.e. corresponding to the eigenstate $|\langle z_1\rangle\rangle$

$$\beth^{+}|\langle z_1\rangle\rangle = [\mathcal{P}_{11}\mathcal{X}_{11} - (\varrho_{11})^2]|\langle z_1\rangle\rangle = \frac{1}{4}|\langle z_1\rangle\rangle \quad (53)$$

The notation $|\langle z_1\rangle\rangle$ is chosen for the state corresponding to the wavefunction in (50) because it can be considered as a generalization of the state $|\langle z\rangle\rangle$corresponding to the wavefunction in (33). The relation (50) is a generalization of (33): (33) is covariant under the action of the particular LCT (2) but (50) is covariant under the action of the general monodimensional LCT (40). The parameters $\mathcal{B}$ in (33) is a real number and is equal to the momentum statistical variances. But the parameter $\mathcal{B}_{11}$ in (50) is a complex number.$\mathcal{B}_{11}$ is related to the coordinate statistical variance and momentum-coordinate covariance by the relation (51). And it can be deduced from (51) and (53) that the relation between $\mathcal{B}_{11}$ and the momentum statistical variance $\mathcal{P}_{11}$ is

$$\mathcal{P}_{11} = \mathcal{B}_{11}(1 + i\frac{\varrho_{11}}{2}) \quad (54)$$

In the limit $\varrho_{11} = 0$, we have as expected $\mathcal{P}_{11} = \mathcal{B}_{11}$.

Given the fact that the operator $\beth^{+}$ in the relation (49) and its lowest eigenvalue are LCT-invariant, it follows that the LCT-transforms of the wavefunction (50) should be of the form

$$\langle x^{1\prime}|\langle z_1\rangle\rangle = e^{iK'}(\frac{1}{2\pi\mathcal{X}'_{11}})^{1/4}e^{-\mathcal{B}'_{11}(x^{1\prime}-\langle x^{1\prime}\rangle)^2 - i\langle p'_1\rangle x^{1\prime}} \quad (55)$$

in which, $e^{iK'}$ is a unitary complex number ($K'$ is real).

### 3.4 Ladder operators and Hamiltonian

Using coordinate representation and the wavefunction in (50) and (55), it can be shown that the state $|\langle z_1\rangle\rangle$ isa common eigenstate of the operator $\boldsymbol{z_1} = \boldsymbol{p_1} + 2i\mathcal{B}_{11}\boldsymbol{x^1}$ andits LCT-transform $\boldsymbol{z'_1} = \boldsymbol{p'_1} + 2i\mathcal{B}'_{11}\boldsymbol{x'^1}$. The corresponding eigenvalue equations are

$$\begin{cases} \boldsymbol{z_1}|\langle z_1\rangle\rangle = \langle z_1\rangle|\langle z_1\rangle\rangle = (\langle p_1\rangle + \frac{2i}{\hbar}\mathcal{B}_{11}\langle x^1\rangle)|\langle z_1\rangle\rangle \\ \boldsymbol{z'_1}|\langle z_1\rangle\rangle = \langle z'_1\rangle|\langle z_1\rangle\rangle = (\langle p'_1\rangle + \frac{2i}{\hbar}\mathcal{B}'_{11}\langle x^{1\prime}\rangle)|\langle z'_1\rangle\rangle \end{cases} \quad (56)$$

we have the “covariant commutation relations”

$$[\boldsymbol{z_1}, \boldsymbol{z_1^\dagger}] = \frac{1}{\mathcal{X}_{11}} \quad [\boldsymbol{z'_1}, \boldsymbol{z_1'^\dagger}] = \frac{1}{\mathcal{X}'_{11}} \quad (57)$$

Then, as generalization of the relation (21), we introduces the following ladder operators $\boldsymbol{z_1}, \boldsymbol{z_1^\dagger}$ andreduced operators $\boldsymbol{p_1}, \boldsymbol{x_1}$ that are covariants under the action of the general monodimensional LCTs (40)

$$
\begin{cases}
\boldsymbol{\mathscr{z}_1} = a_1^1(\boldsymbol{z_1} - \langle z_1 \rangle) = \dfrac{1}{\sqrt{2}}(\boldsymbol{\mathscr{p}_1} + i\boldsymbol{\mathscr{x}_1}) \\
\boldsymbol{\mathscr{z}_1^\dagger} = a_1^{1*}(\boldsymbol{z_1^\dagger} - \langle z_1^* \rangle) = \dfrac{1}{\sqrt{2}}(\boldsymbol{\mathscr{p}}_1^\star - i\boldsymbol{\mathscr{x}}_1^\star)
\end{cases}
\qquad [\boldsymbol{\mathscr{z}_1}, \boldsymbol{\mathscr{z}_1^\dagger}] = 1 \tag{58}
$$

with

$$
\begin{cases}
\boldsymbol{\mathscr{p}_1} = \sqrt{2}a_1^1(\boldsymbol{p_1} - \langle p_1 \rangle) - \sqrt{2}c_1^1(\boldsymbol{x_1} - \langle x_1 \rangle) & \boldsymbol{\mathscr{x}_1} = \sqrt{2}\mathscr{b}_1^1(\boldsymbol{x_1} - \langle x_1 \rangle) \\
\boldsymbol{\mathscr{p}}_1^\star = \sqrt{2}a_1^{1*}(\boldsymbol{p_1} - \langle p_1 \rangle) - \sqrt{2}c_1^{1*}(\boldsymbol{x}_1 - \langle x_1 \rangle) & \boldsymbol{\mathscr{x}}_1^\star = \sqrt{2}\mathscr{b}_1^{1*}(\boldsymbol{x_1} - \langle x_1 \rangle)
\end{cases} \tag{59}
$$

in which the parameters $a_1^1, \mathscr{b}_1^1$ and $c_1^1$ introduced in (58) and (59) are related to the momentum-coordinate variance-covariance by the relation

$$
\begin{cases}
(a_1^1)^2 = \mathcal{X}_1^1 = -\mathcal{X}_{11} \Rightarrow a_1^1 = i\sqrt{\mathcal{X}_{11}} \\
\mathscr{b}_1^1 = \dfrac{1}{2a_1^1} = \dfrac{-i}{2\sqrt{\mathcal{X}_{11}}} \\
c_1^1 = \dfrac{\varrho_{11}}{a_1^1} = \dfrac{-i\varrho_{11}}{\sqrt{\mathcal{X}_{11}}}
\end{cases} \tag{60}
$$

It may be noticed that unlike the operators $\boldsymbol{p_1}$ and $\boldsymbol{x_1}$, the reduced operators $\boldsymbol{\mathscr{p}_1}$ and $\boldsymbol{\mathscr{x}_1}$, in (59) are not, in general, hermitian: in fact, the parameters $a_1^1, \mathscr{b}_1^1$ and $c_1^1$ can have complex values, like in the relation (60), when the metric is not positive definite.

Using the relation (52), the relation (60) can also be put in the matrix form

$$
\begin{pmatrix} \mathcal{P}_{11} & \varrho_{11} \\ \varrho_{11} & \mathcal{X}_{11} \end{pmatrix} = \begin{pmatrix} \mathscr{b}_1^1 & 0 \\ c_1^1 & a_1^1 \end{pmatrix}^T \begin{pmatrix} \eta & 0 \\ 0 & \eta \end{pmatrix} \begin{pmatrix} \mathscr{b}_1^1 & 0 \\ c_1^1 & a_1^1 \end{pmatrix} = \begin{pmatrix} \mathscr{b}_1^1 & 0 \\ c_1^1 & a_1^1 \end{pmatrix}^T \begin{pmatrix} -1 & 0 \\ 0 & -1 \end{pmatrix} \begin{pmatrix} \mathscr{b}_1^1 & 0 \\ c_1^1 & a_1^1 \end{pmatrix} \tag{61}
$$

and we have the properties

$$
\begin{pmatrix} \mathscr{b}_1^1 & 0 \\ c_1^1 & a_1^1 \end{pmatrix}^{-1} = 2\begin{pmatrix} a_1^1 & 0 \\ -c_1^1 & \mathscr{b}_1^1 \end{pmatrix} \tag{62}
$$

From the relations (40), (46), (59), (61) and (62), the law of transformation of the reduced operators $\boldsymbol{\mathscr{p}_1}$ and $\boldsymbol{\mathscr{x}_1}$definedin (61) can be written in the matricial form

$$
(\boldsymbol{\mathscr{p}}_1' \quad \boldsymbol{\mathscr{x}}_1') = (\boldsymbol{\mathscr{p}_1} \quad \boldsymbol{\mathscr{x}_1}) \begin{pmatrix} \Pi_1^1 & \Xi_1^1 \\ \Theta_1^1 & \Lambda_1^1 \end{pmatrix} \tag{63}
$$

with

$$
\begin{pmatrix} \Pi_1^1 & \Xi_1^1 \\ \Theta_1^1 & \Lambda_1^1 \end{pmatrix} = \begin{pmatrix} \Pi_1^1 & -\Theta_1^1 \\ \Theta_1^1 & \Pi_1^1 \end{pmatrix} = 2\begin{pmatrix} \mathscr{b}_1^1 & 0 \\ c_1^1 & a_1^1 \end{pmatrix} \begin{pmatrix} \mathbb{a}_1^1 & \mathbb{c}_1^1 \\ \mathbb{b}_1^1 & \mathbb{d}_1^1 \end{pmatrix} \begin{pmatrix} a_1^{1'} & 0 \\ -c_1^{1'} & \mathscr{b}_1^{1'} \end{pmatrix} \tag{64}
$$

Using the relation (42), the matrix $\begin{pmatrix} \Pi_1^1 & \Xi_1^1 \\ \Theta_1^1 & \Lambda_1^1 \end{pmatrix}$ satisfies the relation

$$\begin{cases} \begin{pmatrix} \Pi_1^1 & \Xi_1^1 \\ \Theta_1^1 & \Lambda_1^1 \end{pmatrix}^T \begin{pmatrix} -1 & 0 \\ 0 & -1 \end{pmatrix} \begin{pmatrix} \Pi_1^1 & \Xi_1^1 \\ \Theta_1^1 & \Lambda_1^1 \end{pmatrix} = \begin{pmatrix} -1 & 0 \\ 0 & -1 \end{pmatrix} \\ \begin{pmatrix} \Pi_1^1 & \Xi_1^1 \\ \Theta_1^1 & \Lambda_1^1 \end{pmatrix}^T \begin{pmatrix} 0 & -1 \\ 1 & 0 \end{pmatrix} \begin{pmatrix} \Pi_1^1 & \Xi_1^1 \\ \Theta_1^1 & \Lambda_1^1 \end{pmatrix} = \begin{pmatrix} 0 & -1 \\ 1 & 0 \end{pmatrix} \end{cases} \quad (65)$$

The relation (65) means that $\begin{pmatrix} \Pi_1^1 & \Xi_1^1 \\ \Theta_1^1 & \Lambda_1^1 \end{pmatrix} = \begin{pmatrix} \Pi_1^1 & -\Theta_1^1 \\ \Theta_1^1 & \Pi_1^1 \end{pmatrix}$ belongs to the group intersection

$$Sp(0,2) \cap O(0,2) \cong SO(0,2) \cong SO(2) \quad (66)$$

From the relations (58) we have for the law of transformation of $\boldsymbol{z_1}$

$$\boldsymbol{z}_1' = \boldsymbol{z_1}\Omega_1^1 = \boldsymbol{z_1}(\Pi_1^1 - i\Theta_1^1) \quad (67)$$

with $\Omega_1^1 = \Pi_1^1 - i\Theta_1^1$ satisfying the relation $\Omega_1^{1*}(-1)\Omega_1^1 = -1$ i.e $\Omega_1^1$ is an element of the group $U(0,1) \cong U(1)$.

From the relations (63), (65) and (67) the following quadratic operator is invariant under the action of the LCT (40) (with $\eta^{11} = \eta_{11} = -1$)

$$\beth^+ = \frac{1}{4}\eta^{11}[(\boldsymbol{p_1})^2 + (\boldsymbol{x_1})^2] = \frac{1}{4}\left(\boldsymbol{z_1}^\dagger\boldsymbol{z_1} + \boldsymbol{z_1}\boldsymbol{z_1}^\dagger\right) = \frac{1}{4}\left(2\boldsymbol{z_1}^\dagger\boldsymbol{z_1} + 1\right) \quad (68)$$

and using the relations (58), (59) and (61) it can be checked that this operator is exactly the quadratic invariant operator in (49). An expression of a general eigenstate, denoted $|n_1, \langle z_1\rangle\rangle$, of the LCT invariant operator can be deduced from the relation (56), (58) and (68)

$$|n_1, \langle z_1\rangle\rangle = \frac{(\boldsymbol{z_1}^\dagger)^{n_1}}{\sqrt{n_1!}}|\langle z_1\rangle\rangle \quad (69)$$

The eigenvalue equation is

$$\beth^+|n_1, \langle z_1\rangle\rangle = \frac{1}{4}(2n_1 + 1)|n_1, \langle z_1\rangle\rangle \quad (70)$$

and we have

$$\begin{cases} [\beth^+, \boldsymbol{z_1}] = -\boldsymbol{z_1} \Rightarrow \boldsymbol{z_1}|n_1, \langle z_1\rangle\rangle = \sqrt{n_1}|n_1 - 1, \langle z_1\rangle\rangle \\ [\beth^+, \boldsymbol{z_1}^\dagger] = \boldsymbol{z_1}^\dagger \ \Rightarrow \ \boldsymbol{z_1}^\dagger|n_1, \langle z_1\rangle\rangle = \sqrt{n_1 + 1}|n_1 + 1, \langle z_1\rangle\rangle \end{cases} \quad (71)$$

The relations (67), (68), (69), (70) and (71) show explicitly that $\boldsymbol{z_1}$ and $\boldsymbol{z_1}^\dagger$ are the ladder operators associated to the LCT-invariant quadratic operator $\beth^+$.

From the relations (58), (59), (60), (61), (62) and (71), we obtain the relation

$$\langle n_1, \langle z_1\rangle|(\boldsymbol{p_1} - \langle p_1\rangle)^2|n_1, \langle z_1\rangle\rangle = (2n_1 + 1)\mathcal{P}_{11} \quad (72)$$

The momentum statistical variance of the state $|n_1, \langle z_1\rangle\rangle$ is equal to $(2n_1 + 1)\mathcal{P}_{11}$ with $\mathcal{P}_{11}$ the statistical variance of the state $|n_1, \langle z_1\rangle\rangle = |\langle z_1\rangle\rangle$ (according to the relation (43)). Given the relations (70) and (72), we may define the general monodimensional momentum dispersion operator $\beth_{11}^{+} = 4\bar{\beth}^{+}\mathcal{P}_{11}$ . The eigenstates of $\beth_{11}^{+}$ are also the states $|n_1, \langle z_1\rangle\rangle$ and the corresponding eigenvalue equation is

$$\beth_{11}^{+}|n_1, \langle z_1\rangle\rangle = 4\bar{\beth}^{+}\mathcal{P}_{11}|n_1, \langle z_1\rangle\rangle = (2n_1 + 1)\mathcal{P}_{11}|n_1, \langle z_1\rangle\rangle \tag{73}$$

The quadratic operator $\bar{\beth}^{+}$ is LCT-invariant but the momentum dispersion operator $\beth_{11}^{+}$ is not invariant because $\mathcal{P}_{11}$ is not invariant (its law of transformation is given by the relation (47)). However, $\beth_{11}^{+}$ is LCT-covariant. Under the action of the LCT (40) we have

$$\beth_{11}' = 4\bar{\beth}^{+}\mathcal{P}_{11}' = \frac{\mathcal{P}_{11}'}{\mathcal{P}_{11}}\beth_{11}^{+} \tag{74}$$

A generalization of the relation (29) that gives an LCT-covariant Hamiltonian $\boldsymbol{H}$ can be deduced from the relation $\beth_{11}^{+} = 4\bar{\beth}^{+}\mathcal{P}_{11}$

$$\boldsymbol{H} = \frac{\langle p_1\rangle^2}{2m} + \frac{\beth_{11}^{+}}{m} = \frac{\langle p\rangle^2}{2m} + \frac{4\bar{\beth}^{+}\mathcal{P}_{11}}{m} = \frac{\langle p_1\rangle^2}{2m} + \frac{\mathcal{P}_{11}}{m}\left(2\boldsymbol{z}_1^{\dagger}\boldsymbol{z}_1 + 1\right) \tag{75}$$

The eigenvalue equation of this LCT-covariant Hamiltonian is

$$\boldsymbol{H}|n_1, \langle z_1\rangle\rangle = [\frac{\langle p_1\rangle^2}{2m} + (2n_1 + 1)\frac{\mathcal{P}_{11}}{m}]|n_1, \langle z_1\rangle\rangle \tag{76}$$

The eigenstates of the LCT-covariant Hamiltonian are also the states $|n_1, \langle z_1\rangle\rangle$.

**Remarks:**

- The set $\{|n_1, \langle z_1\rangle\rangle\}$ (for a fixed $\langle z_1\rangle$) and $\{|\langle z_1\rangle\rangle\}$ are respectively an orthonormal basis and an overcomplete frame corresponding to the space state of the particle. They can be used to define a LCT-covariant phase space representation.
- Instead of the operator $\bar{\beth}^{+}$ in the relation (68), we may consider the simpler quadratic operator

$$\aleph = \aleph_{11} = \boldsymbol{z}_1^{\dagger}\boldsymbol{z}_1 = 2\bar{\beth}^{+} - \frac{1}{2} \tag{77}$$

  which is also an LCT-invariant quadratic operator. $\aleph = \aleph_{11}$ has the property of a number operator of quasi-particles that we may call "dispersion" (because they are related to the momentum dispersion operator). The ladder operators $\boldsymbol{z}_1^{\dagger}$ and $\boldsymbol{z}_1$ are respectively the creation and annihilation operators associated to these quasi-particles. As these quasi-particles are bosons, we may call $\aleph$ a bosonic LCT-invariant quadratic operator (in the section 5, a fermionic LCT- invariant quadratic operator will be also introduced).

## 4- Relation between quantum statistical parameters and thermodynamic variables

Given the relation (30) or (76), it may be asked how to relate a quantum statistical parameter like $\mathcal{B}$ (in (30)) or $\mathcal{P}_{11}$ (in (76)) to quantity that can be measured. In this section, one of our goals is to show that these parameters can be practically related to thermodynamic variables like pressure $P$, volume $V$ and temperature $T$. The formalism of quantum statistical mechanics will be used for this purpose and the explicit example of a simple ideal gas will be considered as illustration. Our study leads also to the introduction of a quantum correction in the thermodynamic state equation of Boltzmann ideal gas.

### 4.1 Hamiltonian operator, Von Neumann entropy and partition function of a particle

We consider the case of a non-relativistic particle in three dimension space at thermal equilibrium with a bath. It corresponds to the canonical ensemble. We denote respectively $P, V$ and $T$ the pressure, the volume allowed to the particle and the temperature. We need to look for the three dimensional generalization of the monodimensional expression (75) of the Hamiltonian $\boldsymbol{H}$. For sake of simplicity, we may suppose that we have the particular case $\mathcal{P}_{11} = \mathcal{P}_{22} = \mathcal{P}_{33} = \mathcal{P}$ (corresponding to an isotropy) and $\mathcal{P}_{ij} = 0$ for $i \neq j$. We may also suppose that the momentum-coordinates covariances $\varrho_{ij}$ are equal to zero then $\mathcal{P} = \mathcal{B}$. We have then for the three dimensional generalization of the expressions (29) or (75) of the Hamiltonian $\boldsymbol{H}$

$$\boldsymbol{H} = \frac{\langle \vec{p} \rangle^2}{2m} + \frac{\beth_{11}^{+} + \beth_{22}^{+} + \beth_{33}^{+}}{m} = \frac{\langle \vec{p} \rangle^2}{2m} + \frac{\mathcal{B}}{m}(2\boldsymbol{z}_1^{\dagger}\boldsymbol{z}_1 + 2\boldsymbol{z}_2^{\dagger}\boldsymbol{z}_2 + 2\boldsymbol{z}_3^{\dagger}\boldsymbol{z}_3 + 3) \tag{78}$$

We may suppose that the particle moves inside the volume $V$ around the center of these volume that one can take as the origin on the coordinates systems so we may choose for the position and momentum global mean values $\langle \vec{x} \rangle = \vec{0}$ and $\langle \vec{p} \rangle = \vec{0}$ and it follows that we have also $\langle z_1 \rangle = 0, \langle z_2 \rangle = 0$ and $\langle z_3 \rangle = 0$. So we finally have for the Hamiltonian

$$\boldsymbol{H} = \frac{\mathcal{B}}{m}(2\boldsymbol{z}_1^{\dagger}\boldsymbol{z}_1 + 2\boldsymbol{z}_2^{\dagger}\boldsymbol{z}_2 + 2\boldsymbol{z}_3^{\dagger}\boldsymbol{z}_3 + 3) \tag{79}$$

The corresponding eigenvalue equation is

$$\boldsymbol{H}|n, \mathcal{B}\rangle = [(2n_1 + 2n_2 + 2n_3 + 3)\frac{\mathcal{B}}{m}]|n, \mathcal{B}\rangle \tag{80}$$

with $|n, \mathcal{B}\rangle = |n_1, n_2, n_3, \langle z_1 \rangle = 0, \langle z_2 \rangle = 0, \langle z_3 \rangle = 0\rangle$ the eigenstate corresponding to the eigenvalue

$$\varepsilon_{n_1,n_2,n_3} = (2n_1 + 2n_2 + 2n_3 + 3)\frac{2\mathcal{B}}{m} \tag{81}$$

The Von Newman entropy of a system described by a density operator $\boldsymbol{\rho}$ is given by the relation

$$S = -kTr[\boldsymbol{\rho} ln \boldsymbol{\rho}] \tag{82}$$

$k$ is the Boltzmann constant and $Tr$ refers to the trace of an operator. At thermal equilibrium, the density operator $\boldsymbol{\rho}$ and the Hamiltonian $\boldsymbol{H}$ commute, $\boldsymbol{\rho H} = \boldsymbol{H \rho}$, and then have the same eigenstates$|n, \mathcal{B}\rangle$.$\boldsymbol{\rho}$ can then be put in the form

$$\boldsymbol{\rho} = \sum_{n_1,n_2,n_3} q_{n_1,n_2,n_3} |n,\mathcal{B}\rangle\langle n,\mathcal{B}| \tag{83}$$

in which $q_{n_1,n_2,n_3}$ are the eigenvalues of $\boldsymbol{\rho}$. Using (83), (82) can be put in the form

$$S = -k \sum_{n_1,n_2,n_3} q_{n_1,n_2,n_3} ln(q_{n_1,n_2,n_3}) \tag{84}$$

Using the maximum entropy principle corresponding to thermal equilibrium, with the constraints ($U = \langle \boldsymbol{H} \rangle$ being the thermodynamical internal energy of the particle)

$$\sum_{n_1,n_2,n_3} q_{n_1,n_2,n_3} = 1 \quad \text{and} \quad U = \langle \boldsymbol{H} \rangle = Tr(\boldsymbol{\rho H}) = \sum_{n_1,n_2,n_3} q_{n_1,n_2,n_3} \varepsilon_{n_1,n_2,n_3} \tag{85}$$

We can deduce the expression of $\boldsymbol{\rho}$ and its eigenvalues $q_{n_1,n_2,n_3}$ (which correspond to a canonical ensemble)

$$q_{n_1,n_2,n_3} = \frac{e^{-\beta \varepsilon_{n_1,n_2,n_3}}}{\mathcal{Z}} \Leftrightarrow \boldsymbol{\rho} = \frac{e^{-\beta \boldsymbol{H}}}{\mathcal{Z}} \tag{86}$$

with $\beta = \frac{1}{kT}$ and $\mathcal{Z}$ the partition function

$$\mathcal{Z} = Tr(e^{-\beta \boldsymbol{H}}) = \sum_{n_1,n_2,n_3} q_{n_1,n_2,n_3} e^{-\beta \varepsilon_{n_1,n_2,n_3}} \tag{87}$$

Using the expression (81) of $\varepsilon_{n_1,n_2,n_3}$, it can be calculated that one has

$$\mathcal{Z} = \sum_{n_1,n_2,n_3} e^{-\beta[(2n_1+2n_2+2n_3+3)\frac{\mathcal{B}}{m}]} = \frac{e^{-3\beta\frac{\mathcal{B}}{m}}}{(1-e^{-2\beta\frac{\mathcal{B}}{m}})^3} = \frac{1}{8sh^3(\beta\frac{\mathcal{B}}{m})} \tag{88}$$

Another way to calculate the partition function $\mathcal{Z}$ utilizes the decomposition of $e^{-\beta \boldsymbol{H}}$ in the overcomplete frame $\{|\langle\vec{z}\rangle\rangle\} = \{|\langle z_1\rangle\rangle, |\langle z_2\rangle\rangle, |\langle z_3\rangle\rangle\}$ which correspond to the phase space representation (generalization of the relation (37))

$$e^{-\beta \boldsymbol{H}} = \int |\langle\vec{z}\rangle\rangle\langle\langle\vec{z}\rangle|e^{-\beta \boldsymbol{H}}|\langle\vec{z}\rangle'\rangle\langle\langle\vec{z}\rangle'| \frac{d^3\langle\vec{p}\rangle d^3\langle\vec{x}\rangle}{h^3}\frac{d^3\langle\vec{p}'\rangle d^3\langle\vec{x}'\rangle}{h^3} \tag{89}$$

The expression of the scalar product $\langle\langle\vec{z}\rangle|n,\mathcal{B}\rangle$ can be used to calculate $\mathcal{Z} = Tr\left(e^{-\beta \boldsymbol{H}}\right)$ from (89) and this calculation gives the same result as in the relation (88)

$$\mathcal{Z} = Tr(e^{-\beta \boldsymbol{H}}) = \int \langle\langle \vec{z}\rangle | e^{-\beta \boldsymbol{H}} | \langle \vec{z}\rangle\rangle \frac{d^3\langle \vec{p}\rangle d^3\langle \vec{x}\rangle}{h^3} = \frac{1}{8sh^3\left(\beta \frac{\mathcal{B}}{m}\right)} \tag{90}$$

The relation between the momentum statistical variance $\mathcal{B}$ and thermodynamic variables can be obtained from the semi-classical limit $\mathcal{B} \to 0$ :

- On one hand, at first order, the expression of $\mathcal{Z}$ in (88) or (90) gives

$$\mathcal{Z} = \frac{1}{8sh^3\left(\beta \frac{\mathcal{B}}{m}\right)} \simeq (\frac{m}{2\beta\mathcal{B}})^3 \tag{91}$$

- On the other hand, the semi-classical approximation of the integral corresponding to (90) is

$$\mathcal{Z} \simeq \frac{1}{h^3}\int e^{-\beta\frac{\vec{p}^2}{2m}} d^3\vec{p} \int_{(V)} d^3\vec{x} = \frac{V}{h^3}(\sqrt{\frac{2\pi m}{\beta}})^3 = \frac{V}{(\lambda_{th})^3} \tag{92}$$

with $\lambda_{th}$ the thermal de Broglie wavelength

$$\lambda_{th} = h\sqrt{\frac{\beta}{2\pi m}} = \frac{h}{\sqrt{2\pi mkT}} \tag{93}$$

From the relations (91) and (92) we obtain the relation

$$\mathcal{B} = \frac{m\lambda_{th}}{2\beta V^{1/3}} = \frac{\hbar^2}{2V^{1/3}\lambda_{th}} = \frac{\hbar\sqrt{2\pi mkT}}{4\pi V^{1/3}} \tag{94}$$

(94) relates the quantum momentum statistical variance $\mathcal{B}$ to thermodynamic.
The expression of the particle partition function (88) using thermodynamic variables is

$$\mathcal{Z} = \frac{1}{8sh^3\left(\beta \frac{\mathcal{B}}{m}\right)} = \frac{1}{8sh^3\left(\frac{\lambda_{th}}{2V^{1/3}}\right)} = \frac{1}{8sh^3\left(\frac{h}{2\sqrt{2\pi mkT}V^{1/3}}\right)} \tag{95}$$

**Remarks**:

- A comparison between the relation (7) and (94) shows that the particle is equivalent to a harmonic oscillator with effective angular frequency

$$\omega = \frac{2\mathcal{B}}{m\hbar} = \frac{\hbar}{mV^{1/3}\lambda_{th}} = \frac{1}{V^{1/3}}\sqrt{\frac{kT}{2\pi m}} \tag{96}$$

- We have called the approximations corresponding to the relations (91) and (92) “semi-classical” but not classical because we didn’t take there $\mathcal{B} = 0$ but only the first order approximation corresponding to the limit $\mathcal{B} \to 0$. The approximated expression (92) of the partition function itself maybe called “semi-classical” because it depends on the thermal de Broglie wavelength which is related to $\mathcal{B}$.

### 4.2 Quantum correction to the Boltzmann ideal gas thermodynamic equation of state

The quantum corrections that are often considered in the study of ideal gas are the corrections related to their bosonic or fermionic nature: these kind of corrections lead respectively to the Bose-Einstein or Fermi-Dirac statistics. However, the quantum correction that we consider here is not of this kind: it is a correction related to the nonzero value of the momentum statistical variance $\mathcal{B}$. Explicitly, this correction corresponds to the fact that the partition function that should be used for a particle is (95) instead of (92). The correction related to the bosonic or fermionic nature of particles is not considered here, it may be considered in future study as extension of the present work.
We consider the ideal gas as a set of $N$ indiscernible particles which correspond to a canonical ensemble. The expression of the corresponding partition function denoted $\mathcal{Z}_N$ is

$$\mathcal{Z}_N = \frac{(\mathcal{Z}_1)^N}{N!} = \frac{(\mathcal{Z})^N}{N!} = \frac{1}{8^N N!\, sh^{3N}\left(\beta \frac{\mathcal{B}}{m}\right)} = \frac{1}{8^N N!\, sh^{3N}\left(\frac{\lambda_{th}}{2V^{1/3}}\right)} \tag{97}$$

with $\mathcal{Z}_1 = \mathcal{Z}$ the one particle partition function given by the relation (95). The thermodynamical free energy $F$ of the gas is given by the relation

$$F = -kTln(\mathcal{Z}_N) \tag{98}$$

and the pressure $P$ is

$$P = -(\frac{\partial F}{\partial V})_{T,N} = (\frac{\partial[kTln(\mathcal{Z}_N)]}{\partial V})_{T,N} \tag{99}$$

The use of the relations (97) and (99) permit to deduce that we have the state equation

$$PV = NkT\left[\beta \frac{\mathcal{B}}{m} coth\left(\beta \frac{\mathcal{B}}{m}\right)\right] = NkT\left[\frac{\lambda_{th}}{2V^{1/3}} coth\left(\frac{\lambda_{th}}{2V^{1/3}}\right)\right] \tag{100}$$

in which $coth$ refers to the hyperbolic cotangent function.
In the semi-classical limits $V \to +\infty$ or $T \to +\infty$ ($\lambda_{th} \longrightarrow 0$) we have

$$\beta \frac{\mathcal{B}}{m} = \frac{\lambda_{th}}{2V^{1/3}} \to 0 \implies \frac{\lambda_{th}}{2V^{1/3}} coth\left(\frac{\lambda_{th}}{2V^{1/3}}\right) \longrightarrow 1 \tag{101}$$

and the ordinary ideal gas equation of state : $PV = NkT$ is obtained.

## 5- Invariant quadratic operators associated to multidimensional general LCTs

### 5.1 Bosonic invariant quadratic operators, reduced operators and ladder operators

Let us consider a $D$-dimensional pseudo-Euclidian space having a metric $\eta$ with signature $(D_+, D_-)$. The momenta and coordinates operators are characterized by the Canonical Commutation Relations (CCRs) [10]

$$\begin{cases} [\boldsymbol{p}_\mu, \boldsymbol{x}_\nu] = \boldsymbol{p}_\mu \boldsymbol{x}_\nu - \boldsymbol{x}_\nu \boldsymbol{p}_\mu = i\eta_{\mu\nu} \\ [\boldsymbol{p}_\mu, \boldsymbol{p}_\nu] = \boldsymbol{p}_\mu \boldsymbol{p}_\nu - \boldsymbol{p}_\nu \boldsymbol{p}_\mu = 0 \\ [\boldsymbol{x}_\mu, \boldsymbol{x}_\nu] = \boldsymbol{x}_\mu \boldsymbol{x}_\nu - \boldsymbol{x}_\nu \boldsymbol{x}_\mu = 0 \end{cases} \quad (102)$$

and the general definition of the LCTs is

$$\begin{cases} \boldsymbol{p}'_\mu = \mathbb{a}_\mu^\nu \boldsymbol{p}_\nu + \mathbb{b}_{\boldsymbol{\mu}}^{\boldsymbol{\nu}} \boldsymbol{x}_\nu \\ \boldsymbol{x}'_\mu = \mathbb{c}_\mu^\nu \boldsymbol{p}_\nu + \mathbb{d}_{\boldsymbol{\mu}}^{\boldsymbol{\nu}} \boldsymbol{x}_\nu \\ [\boldsymbol{p}'_\mu, \boldsymbol{x}'_\nu] = [\boldsymbol{p}_\mu, \boldsymbol{x}_\nu] = i\eta_{\mu\nu} \\ [\boldsymbol{p}'_\mu, \boldsymbol{p}'_\nu] = [\boldsymbol{p}_\mu, \boldsymbol{p}_\nu] = 0 \\ [\boldsymbol{x}'_\mu, \boldsymbol{x}'_\nu] = [\boldsymbol{x}_\mu, \boldsymbol{x}_\nu] = 0 \end{cases} \quad (103)$$

If the $D \times D$ matrices $\mathbb{a}, \mathbb{b}, \mathbb{c}, \mathbb{d}$ and $\eta$ corresponding to the coefficients $\mathbb{a}_\mu^\nu, \mathbb{b}_\mu^\nu, \mathbb{c}_\mu^\nu, \mathbb{d}_\mu^\nu$ and $\eta_{\mu\nu}$ are introduced, the relations in (103) are equivalent to the following matrix relations

$$\begin{cases} \mathbb{a}^T \eta \mathbb{d} - \mathbb{b}^T \eta \mathbb{c} = \eta \\ \mathbb{a}^T \eta \mathbb{b} - \mathbb{b}^T \eta \mathbb{a} = 0 \\ \mathbb{c}^T \eta \mathbb{d} - \mathbb{d}^T \eta \mathbb{c} = 0 \end{cases} \Leftrightarrow \begin{pmatrix} \mathbb{a} & \mathbb{c} \\ \mathbb{b} & \mathbb{d} \end{pmatrix}^T \begin{pmatrix} 0 & \eta \\ -\eta & 0 \end{pmatrix} \begin{pmatrix} \mathbb{a} & \mathbb{c} \\ \mathbb{b} & \mathbb{d} \end{pmatrix} = \begin{pmatrix} 0 & \eta \\ -\eta & 0 \end{pmatrix} \quad (104)$$

According to the relation (104), the $2D \times 2D$ matrix $\begin{pmatrix} \mathbb{a} & \mathbb{c} \\ \mathbb{b} & \mathbb{d} \end{pmatrix}$ belongs to the symplectic group $Sp(2D_+, 2D_-)$[10-11]. The LCT group, that we will denote $\mathbb{T}$ like in the references [10-11] can be then identified with $Sp(2D_+, 2D_-)$.

Let $|\psi\rangle$ be a quantum state. As generalization of the relation (43), one may introduce the following momenta and coordinates statistical mean values and variance-covariances

$$\begin{cases} \langle x_\mu \rangle = \langle \psi | \boldsymbol{x}_\mu | \psi \rangle = \eta_{\mu\nu} \langle \psi | \boldsymbol{x}^\nu | \psi \rangle = \eta_{\mu\nu} \langle x^\nu \rangle \\ \langle p_\mu \rangle = \langle \psi | \boldsymbol{p}_{\boldsymbol{\mu}} | \psi \rangle = \eta_{\mu\nu} \langle \psi | \boldsymbol{p}^{\boldsymbol{\nu}} | \psi \rangle = \eta_{\mu\nu} \langle p^\nu \rangle \\ \mathcal{X}_{\mu\nu} = \langle \psi | (\boldsymbol{x}_{\boldsymbol{\mu}} - \langle x_\mu \rangle)(\boldsymbol{x}_{\boldsymbol{\nu}} - \langle x_\nu \rangle) | \psi \rangle = \eta_{\mu\rho} \mathcal{X}_\nu^\rho \\ \mathcal{P}_{\mu\nu} = \langle \psi | (\boldsymbol{p}_{\boldsymbol{\mu}} - \langle p_\mu \rangle)(\boldsymbol{p}_{\boldsymbol{\nu}} - \langle p_\nu \rangle) | \psi \rangle = \eta_{\mu\rho} \mathcal{P}_\nu^\rho \\ \varrho_{\mu\nu}^{\ltimes} = \langle \psi | (\boldsymbol{p}_{\boldsymbol{\mu}} - \langle p_\mu \rangle)(\boldsymbol{x}_{\boldsymbol{\nu}} - \langle x_\nu \rangle) | \psi \rangle \\ \varrho_{\mu\nu}^{\rtimes} = \langle \psi | (\boldsymbol{x}_{\boldsymbol{\nu}} - \langle x_\nu \rangle)(\boldsymbol{p}_{\boldsymbol{\mu}} - \langle p_\mu \rangle) | \psi \rangle \\ \varrho_{\mu\nu} = \frac{1}{2} \left( \varrho_{\mu\nu}^{\ltimes} + \varrho_{\mu\nu}^{\rtimes} \right) \end{cases} \quad (105)$$

From the commutation relation $[\boldsymbol{p}_\mu, \boldsymbol{x}_\nu] = i\eta_{\mu\nu}$ we obtain the relations

$$\varrho_{\mu\nu}^{\ltimes} - \varrho_{\mu\nu}^{\rtimes} = i\eta_{\mu\nu} \Rightarrow \begin{cases} \varrho_{\mu\nu}^{\ltimes} = \varrho_{\mu\nu}^{\rtimes} + i\eta_{\mu\nu} = \varrho_{\mu\nu} + \frac{i}{2}\eta_{\mu\nu} \\ \varrho_{\mu\nu}^{\rtimes} = \varrho_{\mu\nu}^{\ltimes} - i\eta_{\mu\nu} = \varrho_{\mu\nu} - \frac{i}{2}\eta_{\mu\nu} \\ \varrho_{\mu\nu} = \frac{1}{2} \left( \varrho_{\mu\nu}^{\ltimes} + \varrho_{\mu\nu}^{\rtimes} \right) = \varrho_{\mu\nu}^{\ltimes} - \frac{i}{2}\eta_{\mu\nu} = \varrho_{\mu\nu}^{\rtimes} + \frac{i}{2}\eta_{\mu\nu} \end{cases} \quad (106)$$

Like in the references [10-11], the LCT (103) is considered to be a change of observational frame of reference. Then it is the momenta and coordinates operators which change but not the state $|\psi\rangle$.
Under the action of the LCT (40), the mean values and statistical variance-covariance become

$$\begin{cases} \langle x'_\mu\rangle = \langle\psi|\boldsymbol{x'_\mu}|\psi\rangle = \eta_{\mu\nu}\langle\psi|\boldsymbol{x^{\nu\prime}}|\psi\rangle = \eta_{\mu\nu}\langle x'^{\nu}\rangle \\ \langle p'_\mu\rangle = \langle\psi|\boldsymbol{p'_\mu}|\psi\rangle = \eta_{\mu\nu}\langle\psi|\boldsymbol{p^{\nu\prime}}|\psi\rangle = \eta_{\mu\nu}\langle p^{\nu\prime}\rangle \\ \mathcal{X}'_{\mu\nu} = \langle\psi|(\boldsymbol{x'_\mu} - \langle x'_\mu\rangle)(\boldsymbol{x'_\nu} - \langle x'_\nu\rangle)|\psi\rangle = \eta_{\mu\rho}\mathcal{X}'^{\rho}_{\nu} \\ \mathcal{P}'_{\mu\nu} = \langle\psi|(\boldsymbol{p'_\mu} - \langle p'_\mu\rangle)(\boldsymbol{p'_\nu} - \langle p'_\nu\rangle)|\psi\rangle = \eta_{\mu\rho}\mathcal{P}'^{\rho}_{\nu} \\ \varrho^{\ltimes\prime}_{\mu\nu} = \langle\psi|(\boldsymbol{p'_\mu} - \langle p'_\mu\rangle)(\boldsymbol{x'_\nu} - \langle x'_\nu\rangle)|\psi\rangle \\ \varrho^{\rtimes\prime}_{\mu\nu} = \langle\psi|(\boldsymbol{x'_\nu} - \langle x'_\nu\rangle)(\boldsymbol{p'_\mu} - \langle p'_\mu\rangle)|\psi\rangle \\ \varrho'_{\mu\nu} = \frac{1}{2}\left(\varrho^{\ltimes\prime}_{\mu\nu} + \varrho^{\rtimes\prime}_{\mu\nu}\right) \end{cases} \tag{107}$$

Let us denote respectively $\langle x\rangle, \langle p\rangle, \mathcal{P}, \mathcal{X}$ and $\varrho$ the $1 \times N$ and $N \times N$ matrices corresponding respectively to the mean values and statistical variance-covariances defined in the relations (105). From the relations (103),(105),(106) and (107) we have the law of transformations

$$(\langle p'\rangle \quad \langle x'\rangle) = (\langle p\rangle \quad \langle x\rangle)\begin{pmatrix} \mathbb{a} & \mathbb{c} \\ \mathbb{b} & \mathbb{d} \end{pmatrix} \tag{108}$$

$$\begin{pmatrix} \mathcal{P}' & \varrho' \\ \varrho'^T & \mathcal{X}' \end{pmatrix} = \begin{pmatrix} \mathbb{a} & \mathbb{c} \\ \mathbb{b} & \mathbb{d} \end{pmatrix}^T \begin{pmatrix} \mathcal{P} & \varrho \\ \varrho^T & \mathcal{X} \end{pmatrix}\begin{pmatrix} \mathbb{a} & \mathbb{c} \\ \mathbb{b} & \mathbb{d} \end{pmatrix} \tag{109}$$

As generalization of the relation (49), and from the relations (103), (108) and (109), the general LCT-invariant quadratic operator is

$$\beth^+ = \frac{1}{8}((\boldsymbol{p} - \langle p\rangle) \quad (\boldsymbol{x} - \langle x\rangle))\begin{pmatrix} \mathcal{P} & \varrho \\ \varrho^T & \mathcal{X} \end{pmatrix}^{-1}((\boldsymbol{p} - \langle p\rangle) \quad (\boldsymbol{x} - \langle x\rangle))^T \tag{110}$$

The lowest eigenvalue of $\beth^+$ is equal to $\frac{D}{4}$ and the corresponding eigenstate, denoted $|\langle z\rangle\rangle = |\{\langle z_\mu\rangle\}\rangle$ , is the state corresponding to the coordinate wavefunction of the form [10]

$$\langle\{x^\mu\}|\{\langle z_\mu\rangle\}\rangle = \langle x|\langle z\rangle\rangle = e^{iK}\frac{e^{-\mathcal{B}_{\mu\nu}(x^\mu - \langle x^\mu\rangle)(x^\nu - \langle x^\nu\rangle) - i\langle p_\mu\rangle x^\mu}}{[(2\pi)^D(det\mathcal{X})]^{1/4}} \tag{111}$$

in which $e^{iK}$ is a unitary complex number that doesn't depend on $x^\mu$ . The statistical parameters in the wavefunction (111) corresponds to the state $|\langle z\rangle\rangle$ itself. $\mathcal{B}_{\mu\nu}$ are parameters, related to the momentum and coordinates statistical variance-covariances, which are given by the expression

$$\mathcal{B}_{\mu\nu} = \frac{1}{4}(\eta_{\mu\rho} + 2i\langle\varrho_{\mu\rho}\rangle)\widetilde{\mathcal{X}}^{\rho}_{\nu} \tag{112}$$

in which $\widetilde{\mathcal{X}}_{\mu}^{\rho}$ are related to $\mathcal{X}_{\rho\nu}$ by the relation $\widetilde{\mathcal{X}}_{\mu}^{\rho}\mathcal{X}_{\rho\nu} = \eta_{\mu\nu}$. We have between $\mathcal{P}_{\mu\nu}, \varrho_{\mu\nu}$ and $\widetilde{\mathcal{X}}_{\mu\nu}$ the relation

$$\mathcal{P}_{\mu\nu} = \frac{1}{4}\widetilde{\mathcal{X}}_{\mu\nu} + \varrho_{\mu\alpha}\widetilde{\mathcal{X}}^{\alpha\beta}\varrho_{\nu\beta} \tag{113}$$

The expression of LCT-transform of the wavefunction (111) can also be put in the form [10]

$$\langle\{x'^{\mu}\}|\{\langle z_{\mu}\rangle\}\rangle = \langle x'|\langle z\rangle\rangle = e^{iK'}\frac{e^{-\mathcal{B}'_{\mu\nu}(x'^{\mu}-\langle x'^{\mu}\rangle)(x'^{\nu}-\langle x'^{\nu}\rangle)-i\langle p'_{\mu}\rangle x'^{\mu}}}{[(2\pi)^{D}(det\mathcal{X}')]^{1/4}} \tag{114}$$

From (111) and (114), it can be shown that the state $|\langle z\rangle\rangle$ is a common eigenstate of the operators $\boldsymbol{z}_{\mu} = \boldsymbol{p}_{\mu} + 2i\mathcal{B}_{\mu\nu}\boldsymbol{x}^{\boldsymbol{\nu}}$ and their LCT-transforms $\boldsymbol{z}'_{\mu} = \boldsymbol{p}'_{\boldsymbol{\mu}} + 2i\mathcal{B}'_{\mu\nu}\boldsymbol{x}^{\boldsymbol{\nu}\prime}$ [10]

$$\begin{cases}\boldsymbol{z}_{\boldsymbol{\mu}}|\langle z\rangle\rangle = (\langle p_{\mu}\rangle + 2i\mathcal{B}_{\mu\nu}\langle x^{\nu}\rangle)|\langle z\rangle\rangle = \langle z_{\mu}\rangle|\langle z\rangle\rangle \\ \boldsymbol{z}'_{\mu}|\langle z\rangle\rangle = (\langle p'_{\mu}\rangle + 2i\mathcal{B}'_{\mu\nu}\langle x^{\nu\prime}\rangle)|\langle z\rangle\rangle = \langle z'_{\mu}\rangle|\langle z\rangle\rangle\end{cases} \quad \forall\, \mu \tag{115}$$

Now, as generalization of the relation (61), it can be shown [10] that the $2D \times 2D$ statistical variance-covariances matrix $\begin{pmatrix}\mathcal{P} & \varrho \\ \varrho^{T} & \mathcal{X}\end{pmatrix}$ corresponding to the state $|\langle z\rangle\rangle$ can be factorized in the form

$$\begin{pmatrix}\mathcal{P} & \varrho \\ \varrho^{T} & \mathcal{X}\end{pmatrix} = \begin{pmatrix}\mathscr{b} & 0 \\ 2a c \mathscr{b} & a\end{pmatrix}^{T}\begin{pmatrix}\eta & 0 \\ 0 & \eta\end{pmatrix}\begin{pmatrix}\mathscr{b} & 0 \\ 2ac\mathscr{b} & a\end{pmatrix} \tag{116}$$

in which $a, \mathscr{b}$ and $c$ are $D \times D$ matrices satisfying the following properties

$$\begin{cases}a\mathscr{b} = \mathscr{b}a = \frac{1}{2}I_{D} \quad (I_{D}\text{being here the } D\times D \text{ identity matrix}) \\ a^{T} = \eta a \eta \qquad a^{\dagger} = \eta a^{T} = a\eta \iff a_{\mu}^{\nu *} = a^{\mu\nu} \\ \mathscr{b}^{T} = \eta\mathscr{b}\eta \qquad \mathscr{b}^{\dagger} = \eta\mathscr{b} = \mathscr{b}^{T}\eta \iff \mathscr{b}_{\mu}^{\nu *} = \mathscr{b}^{\nu\mu} \\ c^{T} = 2\eta a c \mathscr{b}\eta\end{cases} \tag{117}$$

we have also the properties (generalization of the relation (62))

$$\begin{pmatrix}\mathscr{b} & 0 \\ 2ac\mathscr{b} & a\end{pmatrix}^{-1} = 2\begin{pmatrix}a & 0 \\ -c & \mathscr{b}\end{pmatrix} \tag{118}$$

$$\begin{pmatrix}\mathcal{P} & \varrho \\ \varrho^{T} & \mathcal{X}\end{pmatrix}^{-1} = 4\begin{pmatrix}a & 0 \\ -c & \mathscr{b}\end{pmatrix}\begin{pmatrix}\eta & 0 \\ 0 & \eta\end{pmatrix}\begin{pmatrix}a & 0 \\ -c & \mathscr{b}\end{pmatrix}^{T} = 4\begin{pmatrix}\eta\mathcal{X}\eta & -\eta\varrho^{T}\eta \\ -\eta\varrho\eta & \eta\mathcal{P}\eta\end{pmatrix} \tag{119}$$

Then the expression (110) of the LCT-invariant quadratic operator becomes

$$ℶ^{+} = \frac{1}{2}\left((\boldsymbol{p} - \langle p\rangle) \quad (\boldsymbol{x} - \langle x\rangle)\right)\begin{pmatrix}a & 0 \\ -c & \mathscr{b}\end{pmatrix}\begin{pmatrix}\eta & 0 \\ 0 & \eta\end{pmatrix}\begin{pmatrix}a & 0 \\ -c & \mathscr{b}\end{pmatrix}^{T}\begin{pmatrix}(\boldsymbol{p} - \langle p\rangle)^{\boldsymbol{T}} \\ (\boldsymbol{x} - \langle x\rangle)^{\boldsymbol{T}}\end{pmatrix} \tag{120}$$

If the momentum and coordinates reduced operators is introduced via the matrix relation

$$(\boldsymbol{\mathscr{p}} \quad \boldsymbol{\mathscr{x}}) = \sqrt{2}(\boldsymbol{p} - \langle p \rangle \quad \boldsymbol{x} - \langle x \rangle)\begin{pmatrix} \mathscr{a} & 0 \\ -c & \mathscr{b} \end{pmatrix} \tag{121}$$

the LCT-invariant quadratic operator (120) becomes

$$ב^{+} = \frac{1}{4}(\boldsymbol{\mathscr{p}} \quad \boldsymbol{\mathscr{x}})\begin{pmatrix} \eta & 0 \\ 0 & \eta \end{pmatrix}\begin{pmatrix} \boldsymbol{\mathscr{p}^T} \\ \boldsymbol{\mathscr{x}^T} \end{pmatrix} = \frac{1}{4}\eta^{\mu\nu}\,(\boldsymbol{\mathscr{p}}_\mu \boldsymbol{\mathscr{p}}_\nu + \boldsymbol{\mathscr{x}}_\mu \boldsymbol{\mathscr{x}}_\nu) \tag{122}$$

The following operators can be identified as the ladder operators

$$\begin{cases} \boldsymbol{\mathscr{z}}_\mu = \dfrac{1}{\sqrt{2}}\left(\boldsymbol{\mathscr{p}}_\mu + i\boldsymbol{\mathscr{x}}_\mu\right) = \mathscr{a}_\mu^\nu(\boldsymbol{z_\nu} - \langle z_{\boldsymbol{\nu}} \rangle) \\ \boldsymbol{\mathscr{z}}_\mu^\star = \dfrac{1}{\sqrt{2}}\left(\boldsymbol{\mathscr{p}}_\mu - i\boldsymbol{\mathscr{x}}_\mu\right) = \mathscr{a}_\mu^\nu\left(\boldsymbol{z}_{\boldsymbol{\nu}}^\dagger - \langle z_\nu^* \rangle\right) = \boldsymbol{\mathscr{z}}^{\boldsymbol{\mu}\dagger} \\ \boldsymbol{\mathscr{z}}_\mu^\dagger = \dfrac{1}{\sqrt{2}}\left(\boldsymbol{\mathscr{p}}_\mu^\dagger - i\boldsymbol{\mathscr{x}}_\mu^\dagger\right) = \mathscr{a}_\mu^{\nu *}\left(\boldsymbol{z}_{\boldsymbol{\nu}}^\dagger - \langle z_\nu^* \rangle\right) = \boldsymbol{\mathscr{z}}^{\boldsymbol{\mu}\star} \end{cases} \tag{123}$$

From the relations in (117), (122), (123) and the CCRs (102), we can deduce the relations

$$\begin{cases} ב^{+} = \dfrac{1}{4}\delta^{\mu\nu}\left(\boldsymbol{\mathscr{z}}_\mu \boldsymbol{\mathscr{z}}_\nu^\dagger + \boldsymbol{\mathscr{z}}_\nu^\dagger \boldsymbol{\mathscr{z}}_\mu\right) = \dfrac{1}{4}\delta^{\mu\nu}(2\boldsymbol{\mathscr{z}}_\mu \boldsymbol{\mathscr{z}}_\nu^\dagger + D) \\ \left[\boldsymbol{\mathscr{z}}_\mu, \boldsymbol{\mathscr{z}}_\nu^\dagger\right] = \delta_{\mu\nu} \quad \left[ב^{+}, \boldsymbol{\mathscr{z}}_\mu\right] = -\dfrac{1}{4}\boldsymbol{\mathscr{z}}_\mu \quad \left[ב^{+}, \boldsymbol{\mathscr{z}}_\mu^\dagger\right] = \dfrac{1}{4}\boldsymbol{\mathscr{z}}_\mu^\dagger \end{cases} \tag{124}$$

with

$$\delta^{\mu\nu} = \delta_{\mu\nu} = \begin{cases} 1 \; if \; \mu = \nu \\ 0 \; if \mu \neq \nu \end{cases} \tag{125}$$

The relations in (124) show explicitly that $\boldsymbol{\mathscr{z}}_\mu$ and $\boldsymbol{\mathscr{z}}_\mu^\dagger$ are as expected the LCT-covariant ladder operators associated to the LCT-invariant quadratic operator $ב^{+}$. A most general eigenstate of $ב^{+}$, denoted $\left|\{n_\mu\}, \langle z_\mu \rangle\right\rangle = |n, \langle z \rangle\rangle$ is obtained by the relation ($n$ is the set of nonegative integers $n_0, n_1, \dots n_{D-1}$)

$$|n, \langle z \rangle\rangle = [\prod_{\mu=0}^{D-1} \frac{\left(\boldsymbol{\mathscr{z}}_\mu^\dagger\right)^{n_\mu}}{\sqrt{n_\mu!}}]|\langle z \rangle\rangle \tag{126}$$

The corresponding eigenvalue equation is

$$ב^{+}|n, \langle z \rangle\rangle = \frac{1}{4}[\sum_{\mu=0}^{D-1}(2n_\mu + 1)]|n, \langle z \rangle\rangle = [\frac{1}{2}(\sum_{\mu=0}^{D-1} n_\mu) + \frac{D}{4}]|n, \langle z \rangle\rangle \tag{127}$$

As generalization of the operator introduced in the relation (77), we may consider the following operators

$$\begin{cases} \aleph_{\mu\nu} = \dfrac{1}{2}(\boldsymbol{\mathscr{z}}_\mu^\dagger \boldsymbol{\mathscr{z}}_\nu + \boldsymbol{\mathscr{z}}_\nu \boldsymbol{\mathscr{z}}_\mu^\dagger) \\ \aleph = \delta^{\mu\nu}\aleph_{\mu\nu} = \delta^{\mu\nu}\boldsymbol{\mathscr{z}}_\mu^\dagger \boldsymbol{\mathscr{z}}_\nu = 2ב^{+} - \dfrac{D}{2} \end{cases} \tag{128}$$

The operator ℵ, which is a generalization of the operator in (77), is also an LCT-invariant quadratic operator that we may call bosonic LCT-invariant quadratic operator.

### 5.2 Spin representation of LCTs and associated invariant quadratic operators

The law of transformations of the $1\times D$ matrices $\boldsymbol{p}, \boldsymbol{x}$ and $\boldsymbol{z}$ which correspond respectively to the reduced operators $\boldsymbol{p}_\mu, \boldsymbol{x}_\mu$ and lowering operator $\boldsymbol{z}_\mu$ can be written in the form [10]

$$\begin{cases} \boldsymbol{p}'_\mu = \Pi_\mu^\nu \boldsymbol{p}_\nu + \Theta_{\boldsymbol{\mu}}^{\boldsymbol{\nu}} \boldsymbol{x}_\nu \\ \boldsymbol{x}'_\mu = -\Theta_\mu^\nu \boldsymbol{p}_\nu + \Pi_{\boldsymbol{\mu}}^{\boldsymbol{\nu}} \boldsymbol{x}_\nu \end{cases} \Leftrightarrow (\boldsymbol{p}' \quad \boldsymbol{x}') = (\boldsymbol{p} \quad \boldsymbol{x}) \begin{pmatrix} \Pi & -\Theta \\ \Theta & \Pi \end{pmatrix} \tag{129}$$

$$\boldsymbol{z}'_\mu = (\Pi_\mu^\nu - i\Theta_{\boldsymbol{\mu}}^{\boldsymbol{\nu}}) \boldsymbol{z}_\mu \Leftrightarrow \boldsymbol{z}' = \boldsymbol{z}(\Pi - i\Theta) = \boldsymbol{z}\Omega \tag{130}$$

with the $2D \times 2D$ matrix $\begin{pmatrix} \Pi & -\Theta \\ \Theta & \Pi \end{pmatrix}$ belonging to the group intersection [10]

$$\mathbb{G} \cong Sp(2D_+, 2D_-) \cap O(2D_+, 2N_-) \cong Sp(2D_+, 2D_-) \cap SO_0(2D_+, 2D_-) \tag{131}$$

in which $SO_0(2D_+, 2D_-)$ is the identity component of indefinite special orthogonal group $SO(2D_+, 2D_-)$. And the matrix $\Omega = \Pi - i\Theta$ belongs to the pseudo-unitary group $U(D_+, D_-)$. One has the group isomorphism

$$\mathbb{G} \cong Sp(2D_+, 2D_-) \cap SO_0(2D_+, 2D_-) \cong U(D_+, D_-) \tag{132}$$

The topological double cover of $\mathbb{G}$ is a subgroup $\mathbb{S}$ of the spin group $Spin(2D_+, 2D_-)$ which is the double cover of $SO_0(2D_+, 2D_-)$. The relation between $\mathbb{S}$ and $\mathbb{G}$ can be defined with a surjective covering map $\mathcal{u}$ according to the relation

$$\begin{cases} \mathcal{u}: \mathbb{S} \longrightarrow \mathbb{G} \\ \quad \mathcal{S} \longmapsto \mathbb{g} = \begin{pmatrix} \Pi & -\Theta \\ \Theta & \Pi \end{pmatrix} \end{cases} \Leftrightarrow \begin{cases} (\boldsymbol{p}' \quad \boldsymbol{x}') = (\boldsymbol{p} \quad \boldsymbol{x}) \begin{pmatrix} \Pi & -\Theta \\ \Theta & \Pi \end{pmatrix} \\ \qquad \mathbb{z}' = \mathcal{S} \mathbb{z} \mathcal{S}^{-1} \end{cases} \tag{133}$$

with $\mathbb{z}$ the operator

$$\mathbb{z} = \frac{1}{\sqrt{2}} (\boldsymbol{\alpha}^{\boldsymbol{\mu}} \boldsymbol{p}_\mu + \boldsymbol{\beta}^{\boldsymbol{\mu}} \boldsymbol{x}_\mu) \tag{134}$$

in which $\alpha^\mu$ and $\beta^\mu$ are the generators of the Clifford algebra $\mathcal{C}\ell(2D_+, 2D_-)$. They verify the following anticommutation relations

$$\begin{cases} \boldsymbol{\alpha}^{\boldsymbol{\mu}} \boldsymbol{\alpha}^{\boldsymbol{\nu}} + \boldsymbol{\alpha}^{\boldsymbol{\nu}} \boldsymbol{\alpha}^{\boldsymbol{\mu}} = 2\eta^{\mu\nu} \\ \boldsymbol{\beta}^{\boldsymbol{\mu}} \boldsymbol{\beta}^{\boldsymbol{\nu}} + \boldsymbol{\beta}^{\boldsymbol{\nu}} \boldsymbol{\beta}^{\boldsymbol{\mu}} = 2\eta^{\mu\nu} \\ \boldsymbol{\alpha}^{\boldsymbol{\mu}} \boldsymbol{\beta}^{\boldsymbol{\nu}} + \boldsymbol{\beta}^{\boldsymbol{\nu}} \boldsymbol{\alpha}^{\boldsymbol{\mu}} = 0 \end{cases} \tag{135}$$

As we have $(\boldsymbol{\alpha}^{\boldsymbol{\mu}})^2 = (\boldsymbol{\beta}^{\boldsymbol{\mu}})^2 = \eta^{\mu\mu} = \pm 1$, we may add the following properties

$$\begin{cases} \boldsymbol{\alpha}^{\boldsymbol{\mu}\dagger} = \boldsymbol{\alpha}^{\boldsymbol{\mu}} \text{ and } \boldsymbol{\beta}^{\boldsymbol{\mu}\dagger} = \boldsymbol{\beta}^{\boldsymbol{\mu}} \qquad if\ \eta^{\mu\mu} = 1 \\ \boldsymbol{\alpha}^{\boldsymbol{\mu}\dagger} = -\boldsymbol{\alpha}^{\boldsymbol{\mu}} \text{ and } \boldsymbol{\beta}^{\boldsymbol{\mu}\dagger} = -\boldsymbol{\beta}^{\boldsymbol{\mu}} \quad if\ \eta^{\mu\mu} = -1 \end{cases} \tag{136}$$

The relation (133) defines a spin representation of LCTs, the element $\mathcal{S}$ of the group $\mathbb{S}$ corresponding to this spin representation can be the putted in the form

$$\boldsymbol{\mathcal{S}} = e^{\mathbb{s}} \tag{137}$$

with $\mathbb{s}$ an element of the Lie algebra $\mathfrak{s}$ of the Lie group $\mathbb{S}$ :

$$\mathbb{s} = \mathbb{s}_{\mu\nu}\Xi^{\mu\nu} \tag{138}$$

in which the set $\{\boldsymbol{\Xi}^{\boldsymbol{\mu\nu}}\}$ is a basis of the Lie algebra $\mathfrak{s}$ and we have

$$\boldsymbol{\Xi}^{\boldsymbol{\mu\nu}} = \begin{cases} \frac{1}{4}[(\boldsymbol{\alpha}^{\boldsymbol{\mu}}\boldsymbol{\alpha}^{\boldsymbol{\nu}} + \boldsymbol{\beta}^{\boldsymbol{\mu}}\boldsymbol{\beta}^{\boldsymbol{\nu}}) + i(\boldsymbol{\alpha}^{\boldsymbol{\mu}}\boldsymbol{\beta}^{\boldsymbol{\nu}} + \boldsymbol{\alpha}^{\boldsymbol{\nu}}\boldsymbol{\beta}^{\boldsymbol{\mu}})] & for\ \mu \neq \nu \\ \frac{1}{2}i\boldsymbol{\alpha}^{\boldsymbol{\mu}}\boldsymbol{\beta}^{\boldsymbol{\mu}} & for\ \mu = \nu \end{cases} \tag{139}$$

The number of elements of the family $\{\Xi^{\mu\nu}\}$ is equal to $D^2$ : it is the dimension of $\mathfrak{s}$ as vectorial space. If we introduce the operators $\boldsymbol{\zeta}^{\boldsymbol{\mu}}, \boldsymbol{\zeta}^{\boldsymbol{\mu}\star}$ and $\boldsymbol{\zeta}^{\boldsymbol{\mu}\dagger}$ through the relations

$$\begin{cases} \boldsymbol{\zeta}^{\boldsymbol{\mu}} = \frac{1}{2}(\boldsymbol{\alpha}^{\boldsymbol{\mu}} + i\boldsymbol{\beta}^{\boldsymbol{\mu}}) \\ \boldsymbol{\zeta}^{\boldsymbol{\mu}\star} = \frac{1}{2}(\boldsymbol{\alpha}^{\boldsymbol{\mu}} - i\boldsymbol{\beta}^{\boldsymbol{\mu}}) = \boldsymbol{\zeta}_{\boldsymbol{\mu}}^{\dagger} \\ \boldsymbol{\zeta}^{\boldsymbol{\mu}\dagger} = \frac{1}{2}(\boldsymbol{\alpha}^{\boldsymbol{\mu}\dagger} - i\boldsymbol{\beta}^{\boldsymbol{\mu}\dagger}) = \boldsymbol{\zeta}_{\boldsymbol{\mu}}^{\star} \end{cases} \tag{140}$$

the expressions of the $\Xi^{\mu\nu}$ can be written in a simpler form

$$\boldsymbol{\Xi}^{\boldsymbol{\mu\nu}} = \frac{1}{2}(\boldsymbol{\zeta}^{\boldsymbol{\mu}\star}\boldsymbol{\zeta}^{\boldsymbol{\nu}} - \boldsymbol{\zeta}^{\boldsymbol{\nu}}\boldsymbol{\zeta}^{\boldsymbol{\mu}*}) \tag{141}$$

The anticommutation relations corresponding to the operators $\boldsymbol{\zeta}^{\boldsymbol{\mu}}$ and $\boldsymbol{\zeta}^{\boldsymbol{\mu}\dagger} = \boldsymbol{\zeta}_{\boldsymbol{\mu}}^{\star}$ can be deduced from the relations (135) and (140), we obtain

$$\begin{cases} \boldsymbol{\zeta}^{\boldsymbol{\mu}}\boldsymbol{\zeta}^{\boldsymbol{\nu}} + \boldsymbol{\zeta}^{\boldsymbol{\nu}}\boldsymbol{\zeta}^{\boldsymbol{\mu}} = 0 \\ \boldsymbol{\zeta}^{\boldsymbol{\mu}\star}\boldsymbol{\zeta}^{\boldsymbol{\nu}\star} + \boldsymbol{\zeta}^{\boldsymbol{\nu}\star}\boldsymbol{\zeta}^{\boldsymbol{\mu}\star} = 0 \\ \boldsymbol{\zeta}^{\boldsymbol{\mu}}\boldsymbol{\zeta}^{\boldsymbol{\nu}\star} + \boldsymbol{\zeta}^{\boldsymbol{\nu}\star}\boldsymbol{\zeta}^{\boldsymbol{\mu}} = \eta^{\mu\nu} \end{cases} \Leftrightarrow \begin{cases} \boldsymbol{\zeta}^{\boldsymbol{\mu}}\boldsymbol{\zeta}^{\boldsymbol{\nu}} + \boldsymbol{\zeta}^{\boldsymbol{\nu}}\boldsymbol{\zeta}^{\boldsymbol{\mu}} = 0 \\ \boldsymbol{\zeta}^{\boldsymbol{\mu}\dagger}\boldsymbol{\zeta}^{\boldsymbol{\nu}\dagger} + \boldsymbol{\zeta}^{\boldsymbol{\nu}\dagger}\boldsymbol{\zeta}^{\boldsymbol{\mu}\dagger} = 0 \\ \boldsymbol{\zeta}^{\boldsymbol{\mu}}\boldsymbol{\zeta}^{\boldsymbol{\nu}\dagger} + \boldsymbol{\zeta}^{\boldsymbol{\nu}\dagger}\boldsymbol{\zeta}^{\boldsymbol{\mu}} = \delta^{\mu\nu} \end{cases} \tag{142}$$

According to the relation (142), the operators $\boldsymbol{\zeta}^{\boldsymbol{\mu}}$ and $\boldsymbol{\zeta}^{\boldsymbol{\mu}\dagger} = \boldsymbol{\zeta}_{\boldsymbol{\mu}}^{\star}$ have the properties of fermionic ladder operators (while the operators $\mathscr{z}_{\mu}$ and $\mathscr{z}_{\mu}^{\dagger}$ in (124) have the properties of bosonic ladder operators). We may introduce a quadratic operator $\boldsymbol{\Sigma}$ defined by the following relation (with a summation on $\mu$)

$$\begin{cases} \boldsymbol{\Sigma}^{\boldsymbol{\mu\nu}} = \boldsymbol{\zeta}^{\boldsymbol{\mu}\dagger}\boldsymbol{\zeta}^{\boldsymbol{\nu}} \\ \boldsymbol{\Sigma} = \boldsymbol{\delta}_{\boldsymbol{\mu\nu}}\boldsymbol{\Sigma}^{\boldsymbol{\mu\nu}} = \boldsymbol{\delta}_{\boldsymbol{\mu\nu}}\boldsymbol{\zeta}^{\boldsymbol{\mu}\dagger}\boldsymbol{\zeta}^{\boldsymbol{\nu}} \end{cases} \tag{143}$$

we have the commutation relation

$$\begin{cases}[\boldsymbol{\Sigma},\boldsymbol{\zeta}^{\boldsymbol{\mu}}]=-\boldsymbol{\zeta}^{\boldsymbol{\mu}}\\ [\boldsymbol{\Sigma},\boldsymbol{\zeta}^{\boldsymbol{\mu}\dagger}]=\boldsymbol{\zeta}^{\boldsymbol{\mu}\dagger}\end{cases} \tag{144}$$

Using the relations (141), (142) and (143), it can be checked that $\boldsymbol{\Sigma}$ commutes with the generators $\Xi^{\mu\nu}$ of the Lie algebra $\mathfrak{s}$ **.** It follows from (137) and (138) that $\boldsymbol{\Sigma}$ commutes with any element $\mathcal{S}$ of the group $\mathbb{S}$ which corresponds to the spin representation of LCTs. In other words, $\boldsymbol{\Sigma}$ is an LCT-invariant quadratic operator. Given the relations (142), (143) and (144), we may call it the fermionic LCT-invariant quadratic operator (like we have called the operator $\aleph$ in (128) a bosonic LCT-invariant quadratic operator.

Now, using the relation (123), (124), (134), (140) and (142), it can be deduced that we have the relation

$$(\not{\boldsymbol{z}})^2=\left(\delta^{\mu\nu}\boldsymbol{z}_\mu^\dagger\boldsymbol{z}_\nu\ +\delta_{\mu\nu}\boldsymbol{\zeta}^{\boldsymbol{\mu}\dagger}\boldsymbol{\zeta}^{\boldsymbol{\nu}}\right)=\aleph+\boldsymbol{\Sigma} \tag{145}$$

in which $\aleph$ is the bosonic LCT-invariant quadratic operator in (131) and $\boldsymbol{\Sigma}$ the fermionic LCT-invariant quadratic operator in (143). The operator $(\not{\boldsymbol{z}})^2$ itself is then an LCT- invariant quadratic operator that we may call the mixed (bosonic-fermionic) LCT- invariant quadratic operator.

### 5.3 Eigenstates of the invariant quadratic operators

Let us denote

$$|n,\not{f},\langle z\rangle\rangle=\left|\{n_\mu\},\{\not{f}^\mu\},\{\langle z_\mu\rangle\}\right\rangle \tag{146}$$

the common eigeinstates of the number-like operators $\aleph_{\mu\mu}=\boldsymbol{z}_\mu^\dagger\boldsymbol{z}_\mu$ and $\boldsymbol{\Sigma}^{\boldsymbol{\mu\mu}}=\boldsymbol{\zeta}^{\boldsymbol{\mu}\dagger}\boldsymbol{\zeta}^{\boldsymbol{\mu}}$ with the eigenvalue equations

$$\begin{cases}\aleph_{\mu\mu}|n,\not{f},\langle z\rangle\rangle=n_\mu|n,\not{f},\langle z\rangle\rangle\\ \boldsymbol{\Sigma}^{\boldsymbol{\mu\mu}}|n,\not{f},\langle z\rangle\rangle=\not{f}^\mu|n,\not{f},\langle z\rangle\rangle\end{cases} \tag{147}$$

and which satisfy the relation

$$\left\langle n,\not{f},\langle z\rangle\middle|\boldsymbol{z}_\mu\middle|n,\not{f},\langle z\rangle\right\rangle=\langle z_\mu\rangle \tag{148}$$

In the relation (146) $n$ refers to the set of the parameters $n_\mu$, $\not{f}$ to the set of the parameters $\not{f}^\mu$ and $\langle z\rangle$ to the set of the parameters $z_\mu$ ($\mu=0,1,\dots,D-1$ ) with $D$ the dimension of the pseudo-Euclidian space that is considered. We may introduce the following quantities

$$\begin{cases}|n|=n_0+n_1+\cdots n_{D-1}\\ |\not{f}|=\not{f}^0+\not{f}^1\dots+\not{f}^{D-1}\end{cases} \tag{149}$$

with this relation (149), the eigenvalue equations of the LCT-invariant quadratic operators $\aleph,\boldsymbol{\Sigma}$ and $(\not{\boldsymbol{z}})^2$implicated in the relation (145) can be written in the compact forms

$$\begin{cases}\aleph|n,\not{f},\langle z\rangle\rangle=|n||n,\not{f},\langle z\rangle\rangle\\ \boldsymbol{\Sigma}|n,\not{f},\langle z\rangle\rangle=|\not{f}||n,\not{f},\langle z\rangle\rangle\\ (\not{\boldsymbol{z}})^2|n,\not{f},\langle z\rangle\rangle=(|n|+|\not{f}|)|n,\not{f},\langle z\rangle\rangle\end{cases} \tag{150}$$

The degeneracy $g_{|n|}$ of $|n|$and $g_{|\not{f}|}$ of $|\not{f}|$are respectively :

$$\begin{cases} g_{|n|} = \frac{(|n| + D - 1)!}{|n|!\,(D-1)!} \\ g_{|\not{f}|} = \frac{D!}{|\not{f}|!\,(D - |\not{f}|)!} \end{cases} \tag{151}$$

The states $|0, \not{f}, \langle z \rangle\rangle = |\not{f}, \langle z \rangle\rangle (|n| = 0)$ are also eigenstates of the operators $\boldsymbol{z}_\mu$

$$\boldsymbol{z}_\mu |\not{f}, \langle z \rangle\rangle = \langle z_\mu \rangle |\not{f}, \langle z \rangle\rangle \tag{152}$$

Any state$|n, \not{f}, \langle z \rangle\rangle$ can be deduced from the state $|0,0, \langle z \rangle\rangle$(($|n| = 0$ and $|\not{f}| = 0$ ) or from any other state using the ladder operators $\boldsymbol{z}_\mu, \boldsymbol{z}_\mu^\dagger, \boldsymbol{\zeta}^{\mu\dagger}$and $\boldsymbol{\zeta}^{\mu}$.

### 5.4 Example of application in Particle Physics

The idea of using LCT spin representation to obtain a classification of elementary fermions has been developed in the references [10-11]. The space then considered is a pentadimensional pseudo-Euclidian space with signature (1,4). Based on the results described in these references, we can establish that a basic quantum state of an elementary fermions can be described with a state $|n, \not{f}, \langle z \rangle\rangle$. As in [10-11], let us consider the following operators:

$$\boldsymbol{\mathcal{Y}}^0 = \frac{1}{2} i\boldsymbol{\alpha}^0\boldsymbol{\beta}^0 \quad \boldsymbol{\mathcal{Y}}^1 = \frac{1}{3} i\boldsymbol{\alpha}^1\boldsymbol{\beta}^1 \quad \boldsymbol{\mathcal{Y}}^2 = \frac{1}{3} i\boldsymbol{\alpha}^2\boldsymbol{\beta}^2 \quad \boldsymbol{\mathcal{Y}}^3 = \frac{1}{3} i\boldsymbol{\alpha}^3\boldsymbol{\beta}^3 \quad \boldsymbol{\mathcal{Y}}^4 = \frac{1}{2} i\boldsymbol{\alpha}^4\boldsymbol{\beta}^4 \tag{153}$$

in which the $\boldsymbol{\alpha}^{\boldsymbol{\mu}}$ and$\boldsymbol{\beta}^{\boldsymbol{\mu}}$ are the generators of the Clifford algebra $\mathcal{C\ell}(2,8)$. They verify anticommutation relations similar to (135). As shown in [10-11], the operators $\boldsymbol{I_3}, \boldsymbol{Y_W}$and $\boldsymbol{Q}$ corresponding respectively to the weak isospin, weak hypercharge and electric charge of an elementary fermion of the Standard Model can be defined from the operators in (153) by the relations

$$\begin{cases} \boldsymbol{I_3} = \frac{1}{2}\boldsymbol{\mathcal{Y}}^0 - \frac{1}{2}\boldsymbol{\mathcal{Y}}^4 \quad \boldsymbol{Y_W} = \boldsymbol{\mathcal{Y}}^0 + \boldsymbol{\mathcal{Y}}^1 + \boldsymbol{\mathcal{Y}}^2 + \boldsymbol{\mathcal{Y}}^3 + \boldsymbol{\mathcal{Y}}^4 \\ \boldsymbol{Q} = \boldsymbol{\mathcal{Y}}^0 + \frac{1}{2}\boldsymbol{\mathcal{Y}}^1 + \frac{1}{2}\boldsymbol{\mathcal{Y}}^2 + \frac{1}{2}\boldsymbol{\mathcal{Y}}^3 = \boldsymbol{I_3} + \frac{\boldsymbol{Y_W}}{2} \end{cases} \tag{154}$$

From the relations (140), we can deduce the relations

$$\begin{cases} \boldsymbol{I_3} = \frac{1}{2}(\boldsymbol{\Sigma}^{\boldsymbol{00}} + \boldsymbol{\Sigma}^{\boldsymbol{44}}) - \frac{1}{2} \\ \boldsymbol{Y_W} = \boldsymbol{\Sigma}^{\boldsymbol{00}} - \frac{2}{3}(\boldsymbol{\Sigma}^{\boldsymbol{11}} + \boldsymbol{\Sigma}^{22} + \boldsymbol{\Sigma}^{33}) - \boldsymbol{\Sigma}^{\boldsymbol{44}} + 1 \\ \boldsymbol{Q} = \boldsymbol{\Sigma}^{\boldsymbol{00}} - \frac{1}{3}(\boldsymbol{\Sigma}^{\boldsymbol{11}} + \boldsymbol{\Sigma}^{22} + \boldsymbol{\Sigma}^{33}) = \boldsymbol{I_3} + \frac{\boldsymbol{Y_W}}{2} \end{cases} \tag{155}$$

It follows from the relation (147) and (155) that the eigenstates of $\boldsymbol{I_3}, \boldsymbol{Y_W}$ and $\boldsymbol{Q}$ are the states $|n, \not{f}, \langle z \rangle\rangle$. The table 1 below gives a classification of the states $|n, \not{f}, \langle z \rangle\rangle$ which corresponds to a classification of a family of elementary fermions, for a fixed value of $n$, and according to the values of the eigenvalues $\not{f}^0, \not{f}^1, \not{f}^2, \not{f}^3$ and $\not{f}^4$ of the operators $\boldsymbol{\Sigma}^{\boldsymbol{\mu\mu}}$ and the eigenvalues $|\not{f}|, I_3, Y_W$ and $Q$ of the operators $\boldsymbol{\Sigma}, \boldsymbol{I_3}, \boldsymbol{Y_W}$ and $\boldsymbol{Q}$. It follows from (147) and (155) that

$$\begin{cases} I_3 = \frac{1}{2}(\not{f}^0 + \not{f}^4) - \frac{1}{2} \\ Y_W = \not{f}^0 - \frac{2}{3}(\not{f}^1 + \not{f}^2 + \not{f}^3) - \not{f}^4 + 1 \\ Q = \not{f}^0 - \frac{1}{3}(\not{f}^1 + \not{f}^2 + \not{f}^3) = I_3 + \frac{Y_W}{2} \end{cases} \quad (156)$$

| **EIGENVALUES OF $\Sigma^{00}, \Sigma^{11}, \Sigma^{22}, \Sigma^{33}, \Sigma^{44}, \Sigma, I_3, Y_W$ AND $Q$** | | | | | | | | | **STATES** |
|---|---|---|---|---|---|---|---|---|---|
| $\not{f}^0$ | $\not{f}^1$ | $\not{f}^2$ | $\not{f}^3$ | $\not{f}^4$ | $\lvert\not{f}\rvert$ | $I_3$ | $Y_W$ | $Q$ | |
| 0 | 0 | 0 | 0 | 0 | 0 | $-1/2$ | 1 | 0 | $\lvert n, \bar{\nu}_L, \langle z \rangle\rangle$ |
| 0 | 0 | 0 | 0 | 1 | 1 | 0 | 0 | 0 | $\lvert n, \nu_R, \langle z \rangle\rangle$ |
| 0 | 0 | 0 | 1 | 0 | 1 | $-1/2$ | $1/3$ | $-1/3$ | $\lvert n, d_L^r, \langle z \rangle\rangle$ |
| 0 | 0 | 0 | 1 | 1 | 2 | 0 | $-2/3$ | $-1/3$ | $\lvert n, d_R^r, \langle z \rangle\rangle$ |
| 0 | 0 | 1 | 0 | 0 | 1 | $-1/2$ | $1/3$ | $-1/3$ | $\lvert n, d_L^g, \langle z \rangle\rangle$ |
| 0 | 0 | 1 | 0 | 1 | 2 | 0 | $-2/3$ | $-1/3$ | $\lvert n, d_R^g, \langle z \rangle\rangle$ |
| 0 | 0 | 1 | 1 | 0 | 2 | $-1/2$ | $-1/3$ | $-2/3$ | $\lvert n, \bar{u}_L^b, \langle z \rangle\rangle$ |
| 0 | 0 | 1 | 1 | 1 | 3 | 0 | $-4/3$ | $-2/3$ | $\lvert n, \bar{u}_R^b, \langle z \rangle\rangle$ |
| 0 | 1 | 0 | 0 | 0 | 1 | $-1/2$ | $1/3$ | $-1/3$ | $\lvert n, d_L^b, \langle z \rangle\rangle$ |
| 0 | 1 | 0 | 0 | 1 | 2 | 0 | $-2/3$ | $-1/3$ | $\lvert n, d_R^b, \langle z \rangle\rangle$ |
| 0 | 1 | 0 | 1 | 0 | 2 | $-1/2$ | $-1/3$ | $-2/3$ | $\lvert n, \bar{u}_L^g, \langle z \rangle\rangle$ |
| 0 | 1 | 0 | 1 | 1 | 3 | 0 | $-4/3$ | $-2/3$ | $\lvert n, \bar{u}_R^g, \langle z \rangle\rangle$ |
| 0 | 1 | 1 | 0 | 0 | 2 | $-1/2$ | $-1/3$ | $-2/3$ | $\lvert n, \bar{u}_L^r, \langle z \rangle\rangle$ |
| 0 | 1 | 1 | 0 | 1 | 3 | 0 | $-4/3$ | $-2/3$ | $\lvert n, \bar{u}_R^r, \langle z \rangle\rangle$ |
| 0 | 1 | 1 | 1 | 0 | 3 | $-1/2$ | $-1$ | $-1$ | $\lvert n, e_L, \langle z \rangle\rangle$ |
| 0 | 1 | 1 | 1 | 1 | 4 | 0 | $-2$ | $-1$ | $\lvert n, e_R, \langle z \rangle\rangle$ |
| 1 | 0 | 0 | 0 | 0 | 1 | 0 | 2 | 1 | $\lvert n, \bar{e}_R, \langle z \rangle\rangle$ |
| 1 | 0 | 0 | 0 | 1 | 2 | $1/2$ | 1 | 1 | $\lvert n, \bar{e}_L, \langle z \rangle\rangle$ |
| 1 | 0 | 0 | 1 | 0 | 2 | 0 | $4/3$ | $2/3$ | $\lvert n, u_R^r, \langle z \rangle\rangle$ |
| 1 | 0 | 0 | 1 | 1 | 3 | $1/2$ | $1/3$ | $2/3$ | $\lvert n, u_L^r, \langle z \rangle\rangle$ |
| 1 | 0 | 1 | 0 | 0 | 2 | 0 | $4/3$ | $2/3$ | $\lvert n, u_R^g, \langle z \rangle\rangle$ |
| 1 | 0 | 1 | 0 | 1 | 3 | $1/2$ | $1/3$ | $2/3$ | $\lvert n, u_L^g, \langle z \rangle\rangle$ |
| 1 | 0 | 1 | 1 | 0 | 3 | 0 | $2/3$ | $1/3$ | $\lvert n, \bar{d}_R^b, \langle z \rangle\rangle$ |
| 1 | 0 | 1 | 1 | 1 | 4 | $1/2$ | $-1/3$ | $1/3$ | $\lvert n, \bar{d}_L^b, \langle z \rangle\rangle$ |
| 1 | 1 | 0 | 0 | 0 | 2 | 0 | $4/3$ | $2/3$ | $\lvert n, u_R^b, \langle z \rangle\rangle$ |
| 1 | 1 | 0 | 0 | 1 | 3 | $1/2$ | $1/3$ | $2/3$ | $\lvert n, u_L^b, \langle z \rangle\rangle$ |
| 1 | 1 | 0 | 1 | 0 | 3 | 0 | $2/3$ | $1/3$ | $\lvert n, \bar{d}_R^g, \langle z \rangle\rangle$ |
| 1 | 1 | 0 | 1 | 1 | 4 | $1/2$ | $-1/3$ | $1/3$ | $\lvert n, \bar{d}_L^g, \langle z \rangle\rangle$ |
| 1 | 1 | 1 | 0 | 0 | 3 | 0 | $2/3$ | $1/3$ | $\lvert n, \bar{d}_R^r, \langle z \rangle\rangle$ |
| 1 | 1 | 1 | 0 | 1 | 4 | $1/2$ | $-1/3$ | $1/3$ | $\lvert n, \bar{d}_L^r, \langle z \rangle\rangle$ |
| 1 | 1 | 1 | 1 | 0 | 4 | 0 | 0 | 0 | $\lvert n, \bar{\nu}_R, \langle z \rangle\rangle$ |
| 1 | 1 | 1 | 1 | 1 | 5 | $1/2$ | $-1$ | 0 | $\lvert n, \nu_L, \langle z \rangle\rangle$ |

The denomination used for a state $\lvert n, \not{f}, \langle z \rangle\rangle$ in the table 1 corresponds to the denomination of the first generation of elementary fermions of the Standard model of particle physics: $\nu$ and $\bar{\nu}$ refer respectively to neutrino and antineutrino, $e$ and $\bar{e}$ refer respectively to negaton and positon, $u$ and $\bar{u}$ refer respectively to up type quark and antiquark, $d$ and $\bar{d}$ refer respectively to down type quark and antiquark. The index $L$ and $R$ correspond to chirality: Left or Right and the exponents $r, g$ and $b$ correspond respectively to strong color charge: red, green and blue. The table 1 correspond to one generation of fermions but it is obtained for a fixed value of $n$ i.e. for a fixed values of the parameters $n_0, n_1, n_2, n_3$ and $n_4$. These parameters may help in the understanding of the existence of multiple generations of fermions. As in [11], the table 1 suggests also the existence of sterile neutrinos.

## 6-Main results

Some of the main results established through this work are the followings:

- There is a relation between the quantum theory of linear harmonic oscillator and Linear Canonical Transformations. According to the relations (2), (3) and (4), the Hamiltonian of a linear harmonic oscillator can be considered as an invariant quadratic operator associated to some particular LCTs.

- According to the relations (7) and (8), the angular frequency of a linear harmonic oscillator can be directly related to the statistical variance of the momentum operator and the Hamiltonian operator can be directly related to the momentum dispersion operator.

- According to the section 3, the theory of a linear harmonic oscillator with nonzero mean values of coordinate and momentum can be exploited to describe the general motion of a non-relativistic quantum particle. This description, which takes into account quantum fluctuation, can be used to obtain a phase space representation of quantum theory.

- According to the section 4, it is possible to relate quantum statistical parameters like momentum statistical variance, which are present in the expression of LCT-invariant quadratic operators, with thermodynamic variables such as temperature, pressure and volume through the formalism of quantum statistical mechanics.

- According to the relations (39), (49) and (68), a first invariant quadratic operator associated to general monodimensional LCTs, denoted $\beth^{+}$ , can be considered as a generalization of a reduced momentum dispersion operator. Another invariant quadratic operator, denoted $\aleph$, which is directly related to $\beth^{+}$ by the relation (77) can be considered as a number operator of bosonic quasiparticles. We may then call $\aleph$ a bosonic invariant quadratic operator. The multidimensional generalization of $\beth^{+}$ and $\aleph$ are considered in the relations (110), (120), (122), (124) and (128). The common eigenstates of these operators which are considered in the relation (70), (126) and (127) can be considered as generalization of the basic state of a linear harmonic oscillator and generalization of what are called coherent state and squeezed state in the literature [13-16].

- Following the relation (142), (143), (144), (145) and (150), a fermionic analog of the bosonic LCT-invariant quadratic operator $\aleph$ can be also considered. This fermionic LCT-invariant quadratic operator is denoted $\boldsymbol{\Sigma}$. The sum $(\not\equiv)^2 = \aleph + \boldsymbol{\Sigma}$ considered in the relations (145) and (150) is itself an LCT-invariant quadratic operator. The eigenvalues equation corresponding to these operators are given in (150).

- The common eigenstates of the LCT-invariant quadratic operators associated to a pentadimensional pseudo-euclidian space with signature (1,4) correspond to basic quantum states of elementary fermions. Following the table 1, a classification of these basic states lead to a classification of elementary fermions which is compatible with the Standard model of particle physics, suggest the existence of sterile neutrino and may help in the understanding of the existence of multiple fermions family.

## 7-Conclusion

The expressions of the general LCT-invariant quadratic operators that were identified are given by the relations (124), (128), (143) and (145). Their eigenvalues equations are given in the relation (150).

The analysis of the results enumerated in the previous section show that the LCT-invariant quadratic operators and their eigenstates can have important roles to play in physics. These results can especially help in the understanding and resolution of some of the main open fundamental problems related to quantum theory, statistical physics, relativistic quantum thermodynamics and particle physics like the question considered in the reference [17], the unsolved questions related to sterile neutrinos [11] and the origin of multiple generations of fermions.

Given the possible link between LCT-covariance and fundamental interactions highlighted in the references [10-11], the results obtained through this work can also be a particular help in the establishment of a unified theory of interaction which includes gravity.